\documentclass[a4paper,10pt]{article}

\usepackage[a4paper,left=2.73cm,right=2.7cm,top=3cm,bottom=3.5cm]{geometry}
\usepackage{amsmath,mathenv,amssymb,subfigure}

\def\be{\begin{equation}}
\def\ee{\end{equation}}
\def\bea{\begin{eqnarray}}
\def\eea{\end{eqnarray}}

\def\tD{\tilde \Delta}
\def\AR{A_{1}}

\def\w{\wedge}
\def\J{J_{KE}}

\def\cA{{\cal A}}
\def\cF{{\cal F}}
\def\customsign{\sigma}

\def\sten{\overset{10}{\star}}
\def\sfive{\overset{5}{\star}}

\begin{document}

\hfill{DFTT 09/2012}

\hfill{ITP-UU-12/28}

\hfill{SPIN-12/26}

\vspace{1cm}

\begin{center}
{\huge{\bf Consistent reduction of charged D3-D7 systems}}
\end{center}

\vskip 10pt

\begin{center}
{\large Aldo L. Cotrone$^{a}$ and Javier Tarr\'\i o$^{b}$}
\end{center}

\vskip 10pt
\begin{center}
\textit{$^a$ Dipartimento di Fisica Teorica, Universit\`a di Torino and I.N.F.N. - sezione di Torino, Via P. Giuria 1, I-10125 Torino, Italy.}\\
\textit{$^b$ Institute for Theoretical Physics and Spinoza Institute, Universiteit Utrecht, 
Leuvenlaan 4, 3584 CE, Utrecht, The Netherlands.}\\

\vskip 10pt
{\small cotrone@to.infn.it, l.j.tarriobarreiro@uu.nl}
\end{center}

\vspace{15pt}

\abstract{
We provide a consistent reduction to five dimensions of the system of D3-branes at Calabi-Yau singularities coupled to D7-branes with world-volume gauge flux. 
The D3-branes source the dual to would-be conformal quiver theories.
The D7-branes, which are homogeneously distributed in their transverse directions, are dual to massless matter in the fundamental representation at finite (baryon) density.
We provide the five-dimensional action and equations of motion, and discuss a few sub-truncations.
The reduction can be used in the study of transport properties and stability of D3-D7 charged systems. 
}

\vfill 

\newpage

\tableofcontents

\section{Introduction}

Consistent reductions of type IIB supergravity on (squashed) Sasaki-Einstein (SE) manifolds have been recently studied in many instances \cite{buchel,Gauntlett:2007ma,Maldacena:2008wh,Gauntlett:2009zw,Gubser:2009qm,Cassani:2010uw,Liu:2010sa,Gauntlett:2010vu,Skenderis:2010vz,Cassani:2010na,Bena:2010pr,Liu:2010pq,Liu:2011dw,Halmagyi:2011yd}.
Apart from their intrinsic supergravity interest, they provide the starting point for the analysis of properties of some dual field theories according to the gauge/gravity correspondence. 
The dual field theories, which include four-dimensional quiver conformal gauge theories, live on the world-volume of $N_c$ D3-branes at the tip of Calabi-Yau cones, whose bases are the Sasaki-Einstein (SE) manifolds.
Notable examples include ${\cal N}=4$ $SU(N_c)$ SYM \cite{Maldacena:1997re} (the dual SE manifold being the five-sphere) and the Klebanov-Witten theory \cite{Klebanov:1998hh} (with the SE manifold being $T^{1,1}$), in the large $N_c$, large 't Hooft coupling limit. 
The gravitational systems always admit an $AdS_5$ solution.
The explicit reductions of the IIB theory to five dimensions allow to study in a consistent setting features such as field theory vacua (i.e. the construction of exact gravity solutions) or transport properties of the dual field theories. 
Moreover, they ease the understanding of the holographic dictionary.

In this paper we are interested in deformations of the above mentioned conformal theories by the inclusion of flavor degrees of freedom.
To this end, we will consider D7-brane embeddings corresponding to massless flavors in the fundamental representation.
Our primary interest is in the case where the world-volume gauge field is turned on, corresponding to the dual field theories being at finite (baryon) charge density (for a discussion of the uncharged case, see \cite{Cassani:2010na}).
This class of theories are benchmark examples to describe holographically systems such as the quark-gluon plasma or condensed matter theories at finite density from a top-down approach. 

The general aim is the study of solutions, transport and stability of this class of D3-D7 systems.
The latter include in particular the D3-D7 plasmas constructed in \cite{Bigazzi:2009bk,Bigazzi:2011it} and reviewed in \cite{Nunez:2010sf,Bigazzi:2011db}; the uncharged case has been further investigated in \cite{Bigazzi:2009tc,Bigazzi:2010ku,Magana:2012kh}, and here we take the first step towards the study of the charged system (the five-dimensional reduction being much more complicated than in the uncharged case).

To this end, we provide the consistent truncation of type IIB supergravity on squashed SE manifolds, coupled to the  Dirac-Born-Infeld (DBI) and a Wess-Zumino (WZ) actions to take into account the D7-brane sources.
Thus, we generalize, in a non-trivial way, the constructions in \cite{Cassani:2010uw,Liu:2010sa,Gauntlett:2010vu,Skenderis:2010vz} to include (a subset of) the D7-brane fields.
Out of the two scalars (the transverse directions to the D7-branes) and the eight components of the gauge field living on the branes, we will retain only a vector field and a scalar (deriving from the gauge field too) in five dimensions.
This corresponds to the minimal choice for having a consistent truncation.
The latter includes also the solutions in \cite{Filev:2011mt,Ammon:2012qs}.

The dual field theories are described by these systems in their Veneziano limit, in which $N_c \rightarrow \infty$ and $N_f/N_c$ is kept fixed, i.e. beyond the ``quenched'' approximation.
This implies that not only the stack of D3-branes affects the dynamics of the D7-branes, but that the latter backreact onto the geometry, modifying the solutions. 
The effects of the backreaction consist in a modification of the volume of the SE manifold, which becomes radial dependent (here and in the following ``radial'' refers to the radius of $AdS$), and its squashing. 
We will choose not to break further symmetries with the D7-branes by working in the smeared limit, where the (many) D7-branes are homogeneously distributed in their transverse directions \cite{Bigazzi:2005md,Casero:2006pt}.
This also allow the use of the Abelian DBI.

The paper is organized as follows.
In section \ref{sec.5deoms} we present the IIB+DBI+WZ ten-dimensional action and the relevant equations of motion to be reduced.
We then set-up the reduction ansatz respecting the isometries of the squashed Sasaki-Einstein compact manifold. 
Section \ref{sec.action} contains the five-dimensional reduced action, while section \ref{sec.eqs} provides all the equations of motion.
We comment on the spectrum of operators of the dual field theories in section \ref{sec.comments}, which also contains three baryon-uncharged sub-truncations; one of them is novel in that it contains a D7 world-volume scalar besides a single massive vector.
We end up with a summary and mention further developments in section \ref{sec.end}.
The appendices contain technical details on the reduction of the brane action and the topological terms.  

{\bf A note on notation:} our notation for the five-dimensional differential forms is of the form $G^{(i)}_{\alpha}$, where the index $\alpha$ denotes the degree of the form in five dimensions, while the index in parenthesis $i=0,\cdots,5$ (whenever present) denotes the degree of the ten-dimensional form reducing to $G^{(i)}_{\alpha}$.   


\section{Setup}\label{sec.5deoms}

At the gravitational level, the presence of space-filling, non-compact D7-branes  is modeled  by the presence of  DBI and WZ terms, which complement the type IIB supergravity action describing the closed string sector. We mostly use the conventions in \cite{Benini:2007kg} which give
\be\label{eq.oldaction}
S=S_{IIB}+S_{sources} \ ,
\ee
where for every D7-brane we have a contribution from a DBI+WZ action
\bea\label{eq.IIBaction}
S_{IIB} & = & \frac{1}{2\kappa_{10}^2} \int \Bigg[  R \sten 1 - \frac{1}{2}d \Phi \w \sten d\Phi - \frac{1}{2}e^{-\Phi} H_{3} \w \sten H_{3}- \frac{1}{2}e^{2\Phi} F_{1} \w \sten F_{1} \\
&&\qquad\qquad - \frac{1}{2}e^{\Phi} F_{3} \w \sten F_{3}- \frac{1}{4} F_{5} \w \sten F_{5} \Bigg]  -\frac{1}{4\kappa_{10}^2}\int C_{4} \w H_{3} \w F_{3}\  , \nonumber \\
S_{sources} & = & \sum^{N_f}\left[ -\frac{1}{2\kappa_{10}^2} \int d^8\xi e^\Phi \sqrt{-\det\left( \hat G + e^{-\Phi/2} \cF \right)} + \frac{\customsign}{2\kappa_{10}^2} \int \hat C_{q} e^{-\cF} \right] \ . \label{eq.sources}
\eea
In these expressions hatted quantities are pulled-back to the worldvolume of the D7-branes, which are extended along the five-dimensional target space of interest and wrap a compact three-cycle in the squashed SE manifold (see section \ref{sec.ansatz} for more details). The RR field strengths are given by $F_{n} = d C_{n-1}-C_{n-3}\w H_{3}$ (the sign is related to the explicit sign in the exponential of the WZ term), and the NSNS field strength is exact $H_{3}=dB_{2}$. The D7-branes admit a gauge flux on their worldvolume, $\cF=2\pi\alpha' d\cA + \hat B_{2}$, being $\cA$ the gauge field. We have included an explicit sign, $\customsign$, in the WZ term to study at the same time D7-branes and anti D7-branes. They do not share modes with the same mass (in other words, they do not give rise to dual operators of the same dimensionality).
Specifically, with our conventions,
\bea
\customsign= \begin{cases} +1 & {\rm for\ anti-branes} \\ -1 & {\rm for\ branes} \end{cases} \ . 
\eea

The source terms involve only the worldvolume coordinates of the D7-branes, and introduce Dirac deltas in the equations of motion of the fields, breaking part of the symmetries of the compactification manifold. This introduces a series of technical problems that can be avoided by performing a smearing of the D7-branes in their transverse directions. This amounts to consider \cite{Bigazzi:2005md,Casero:2006pt,Benini:2006hh}
\be
\sum^{N_f} \int d^8 \xi \, \varphi_{8} \to \int d^{10}x\, \varphi_{8} \wedge \Omega_{2} \ ,
\ee
for every eight-form $\varphi_{8}$ on the D7-branes. Here, $\Omega_{2}$ is a localized form, with at least one component orthogonal to every D7-brane, and an appropriate normalization to extend exactly $N_f$ branes in the transverse directions. This prescription is not straightforward to implement in the DBI part, since the latter is not written in form language; see appendix \ref{app.DBI} for details.

Taking into consideration again \eqref{eq.IIBaction}-\eqref{eq.sources} one can find the equations of motion and (violation of) the Bianchi identities in a straightforward manner \cite{Benini:2007kg}
\bea
\label{BIF1}dF_{1}&=&-\customsign\, \Omega_{2}\ ,\\
dF_{3}&=& H_{3} \wedge F_{1} - \customsign \cF \wedge \Omega_{2} \ ,\\
\label{BIF5}dF_{5}&=& H_{3} \wedge F_{3} - \customsign \frac12 \cF \wedge  \cF \wedge \Omega_{2} \ ,\\
dH_{3}&=&0 \ ,\\ \label{eqforF1}
d\left(e^{\Phi}\sten F_{1}\right)&=&-e^{\Phi}H_{3}\wedge \sten F_{3}- \customsign \frac{1}{24}\cF^4 \wedge \Omega_{2}\ ,\\ \label{eqforF3}
d\left(e^{\Phi}\sten F_{3}\right)&=&-H_{3}\wedge F_{5}+ \customsign \frac{1}{6}\cF^3 \wedge \Omega_{2}\ ,\\ \label{eqforF5}
d\sten F_{5}&=& dF_{5}\ ,\\ \label{eqforH3}
d\left(e^{-\Phi}\sten H_{3}\right)&=& e^{\Phi}F_{1}\wedge \sten F_{3}-F_{5}\wedge F_{3}-2\kappa_{10}^2\frac{\delta S_{sources}}{\delta \cF}\ .
\eea
Of course, these are to be supplemented by the dilaton and Einstein equations
\be\label{eqforP}
d\sten d\Phi = - \frac{1}{2}e^{-\Phi} H_{3} \w \sten H_{3} +  e^{2\Phi} F_{1} \w \sten F_{1} + \frac{1}{2} e^{\Phi} F_{3} \w \sten F_{3} - 2\kappa_{10}^2 \frac{\delta S_{sources}}{\delta \Phi} = 0\ ,
\ee
and
\bea\label{eqforR}
R^{AB} & = &  \frac{1}{2} \partial^A \Phi \partial^B \Phi+ \frac{1}{2}  e^{2 \Phi }   F_{1}^A \lrcorner  F_{1}^B   + \frac{1}{4}  \left( \iota^A F_{5} \right) \lrcorner \left( \iota^B F_{5} \right)   \\
&&  +  \frac{1}{2}  e^{ \Phi } \left[ \left( \iota^A F_{3} \right) \lrcorner \left( \iota^B F_{3} \right) - \frac{1}{4} G^{AB} F_{3} \lrcorner F_{3}  \right]  \nonumber\\
&&  + \frac{1}{2}  e^{- \Phi } \left[ \left( \iota^A H_{3} \right) \lrcorner \left( \iota^B H_{3} \right) - \frac{1}{4} G^{AB} H_{3} \lrcorner H_{3}  \right]   \nonumber \\
&&  + 2\kappa_{10}^2 \left[ \frac{\delta S_{sources} }{\delta G_{AB}} - \frac{1}{8} G^{AB}  G^{CD} \frac{\delta S_{sources}}{\delta G_{CD}} \right]  \nonumber \ ,
\eea
where we are using the conventions in appendix A of \cite{Cassani:2010uw}.

We will be interested in configurations dual to massless flavors.
In this case, the embedding of the branes is trivial.
So, the only active field in the open string sector is the gauge field $\cA$.
Its equations derive from the ones of $H_{3}$ (see appendix \ref{app.NSNS}). 

\subsection{Reduction ansatz}\label{sec.ansatz}

The reduction ansatz for the ten-dimensional metric will have a form
\be\label{eq.ansatzmetric2}
d s_{10}^2 = e^{\frac{10}{3}f} ds_{5}^2 + e^{-2(f+w)} R_{SE}^2 \left[ ds_{KE}^2 + e^{10w} \left( \eta + \AR \right) \otimes  \left( \eta + \AR \right) \right] \ ,
\ee
where $\eta \equiv d\tau + A_{KE}$, and $A_{KE}$ is a one-form depending only on the angles of the compact manifold, whereas $\AR$ can depend on the coordinates of the five-dimensional target space-time. ``$KE$'' denotes a four-dimensional K\"ahler-Einstein (KE) manifold, with an associated K\"ahler form given by $J_{KE} = 2 d A_{KE}$. 
When $w=\AR=0$ we recover the metric of the un-squashed SE manifold, expressed as a $U(1)$ fibration, with fiber given by $\eta$, over the KE base. 
$R_{SE}$ is the radius of the squashed SE manifold, which we will set to $R_{SE}=1$ in the rest of the paper.

From an effective five-dimensional point of view in which the squashed SE manifold is integrated out, $f$ and $w$ become two scalars.
$\AR$ represents, once the reduction over the SE manifold is performed, a $U(1)_R$ gauge field. In principle there exists the possibility to consider charged matter under this gauge field in the reduction, but we will restrict to the neutral case in which these matter fields are not present. Despite considering this truncation, the reduction will still be consistent, although non-supersymmetric.\footnote{Being interested in theories at finite density and/or finite temperature, the lack of supersymmetry is not that relevant after all.}

The reduction ansatz makes heavy use of the structure of the SE manifold.
To begin with, the $F_{5}$ RR field strength satisfies a self-duality condition that cannot be implemented in the action and must be imposed directly in the equations of motion. However, when we perform the reduction on the Sasaki-Einstein manifold we would like to make the self-duality explicit for the dimensionally reduced fields. In order to implement this condition in the reduced equations we find it convenient to consider the following action, adapted to the SE case, 
\bea \label{eq.newaction}
S & = & \frac{1}{2\kappa_{10}^2} \int \Bigg[  R \sten 1 - \frac{1}{2}d \Phi \w \sten d\Phi - \frac{1}{2}e^{-\Phi} H_{3} \w \sten H_{3}- \frac{1}{2}e^{2\Phi} F_{1} \w \sten F_{1} \\
&&\qquad\qquad - \frac{1}{2}e^{\Phi} F_{3} \w \sten F_{3}- \frac{1}{4} F_{5} \w \sten F_{5} \Bigg]  + \frac{1}{2\kappa_{10}^2} \int {\cal L}_{top} + S_{DBI} \ , \nonumber 
\eea
where the field strength definitions are modified to satisfy the Bianchi identities \eqref{BIF1}-\eqref{BIF5}
\bea\label{eq.newdefF1}
F_{1} &=&  dC_{0} +F_{1}^{D7}\ , \\
F_{3} &=& dC_{2} - C_{0} H_{3} + \cF \w F_{1}^{D7} \ , \label{eq.newdefF3} \\
F_{5} &=& F_{fl}^{(5)}+ dC_{4} - \frac{1}{2} C_{2} \w H_{3} + \frac{1}{2} B_{2} \w dC_{2} + \frac{1}{2} \cF\w \cF \w F_{1}^{D7} \ , \label{eq.newdefF5}
\eea
and we have defined\footnote{We use the same notation as \cite{Cassani:2010na}.}
\bea
F_{1}^{D7} \equiv \customsign\, Q_f \left( d\tau +A_{KE}\right)\ .
\eea
Here $F_{fl}^{(5)}$ is the flux part of $F_{5}$, $F_{1}^{D7}$ is the part of $F_{1}$ sourced by the D7-branes and
\be
Q_f \equiv \frac{{V}(X_3)}{4\, {V}(X_{SE})} g_sN_f \ ,
\ee
where ${V}(X)$ is the volume of the manifold $X$, and $X_3$ is the compact three-cycle wrapped by each D7.
The topological term is given by
\bea \label{topo}
{\cal L}_{top} &=& \frac14 \left( F_{fl}^{(5)}+ dC_{4} - \frac{1}{2} \cF^2 \w F_{1}^{D7}\right) \w \left( B_{2} \w dC_{2} - C_{2} \w dB_{2} + \cF^2 \w F_{1}^{D7} \right) \\
&&  - \frac{1}{6} F_{3} \w \cF^3 \w F_{1}^{D7} + \frac{1}{24}F_{1} \w \cF^4 \w F_{1}^{D7}   \ . \nonumber
\eea
This action produces the same equations of motion as the original one  \eqref{eq.IIBaction}-\eqref{eq.sources}, but it has the advantage that the four-form potential $C_{4}$ enters only via derivatives, which makes easier the implementation of the self-duality of $F_{5}$ in the reduced five-dimensional action.

In order to write down the reduction ansatz for the form fields, let us recall some properties of the K\"ahler form and the fiber $\eta$. These satisfy
\be
\J \w \J = 2 {\cal V}(X_{KE}) \ , \quad d\eta = 2 \J \ , \quad  \J \w \J \w \eta = 2{\cal V}(X_{SE}) \ ,
\ee
with ${\cal V}(X)$ the volume form of the manifold $X$. 
We find it convenient to define $n$-forms $\Delta_{n}$ as
\bea
&& \Delta_{0}=1 \ , \quad\Delta_{1} =  \eta \ , \quad \Delta_{2} =  \frac{1}{\sqrt{2}}\J \ , \quad \Delta_{3} =  \frac{1}{\sqrt{2}}\J \w \eta  \ , \\
&&  \Delta_{4} = \frac{1}{2} \J \w \J \ ,  \quad \Delta_{5} = \frac{1}{2}  \J \w \J \w \eta \ , \quad  \Delta_{>5}=0 \ .
\eea
These forms are useful to decompose generic ten-dimensional forms into effective five-dimensional forms, since we can `factor' out the components along the compact squashed SE manifold with the $\Delta_{n}$.
It is useful to remark here that these satisfy
\be
\Delta_{m} \w \Delta_{n} = \begin{cases} 0 & \text{if m and n are odd} \\ \Delta_{m+n} & \text{otherwise} \end{cases} \ ,
\ee
from where  $\Delta_{n}\w\Delta_{5-n} = {\cal V}(X_{SE})$. The Hodge operation with respect to the SE manifold is  implemented trivially, since $*\Delta_{n} = \Delta_{5-n}$.

To include the presence of the $U(1)_R$ gauge field in our ansatz it is better to define a new basis of forms $\tD_{n}$ as follows
\bea
&& \tD_{1} = \Delta_{1} + \AR \ , \qquad \tD_{2} = \Delta_{2}  \ , \qquad \qquad \tD_{3} = \tD_{2} \w \tD_{1}\ ,\\
&& \tD_{4} = \tD_{2} \w \tD_{4}  \ , \qquad  \tD_{5} = \tD_{2} \w \tD_{3}\ . \nonumber
\eea
Some important properties to work with the new basis are
\be
d\tD_{n} = 2\sqrt{2} \left( \tD_{2} \delta_{1n}+\tD_{4} \delta_{3n} \right) +  d\AR \delta_{1n}+\tD_{2} \w d\AR \delta_{3n} \ ,
\ee
and
\be
G_{5} \w \tD_{n} = G_{5} \w \Delta_{n} \ ,
\ee
where $G_{5}$ is any five-form in the target spacetime.

Coming back to the D7-branes, in order to preserve the structure of the SE manifold the pullback of the ten-dimensional spacetime to the worldvolume of the flavor branes gives (considering no gauge field turned on in the worldvolume and a trivial embedding profile)
\be
{\cal V}(D7) = e^{\frac{16}{3}f+2w}\sfive 1 \wedge \Delta_{2} \wedge \Delta_{1} \ ,
\ee
where $\sfive 1$ is the volume form of the target five-dimensional spacetime with metric $ds_5^2$ in \eqref{eq.ansatzmetric2}. In order for $\Omega_{2}$ to be orthogonal to ${\cal V}(D7) $ we need to have $\Omega_{2} \propto \Delta_{2}$. Furthermore, the Bianchi identity for $F_{1}$ is
\be
dF_{1} = -\customsign\, \Omega_{2}  \quad \Rightarrow \quad d\Omega_{2} = 0 \ ,
\ee
and a minimal way to fulfill the requirements is to let $\Omega_{2} = - \customsign Q_f\, d \Delta_{1}$.

Finally, in order to obtain a set of differential equations for five-dimensional quantities we split the different components of the fields into effective five-dimensional forms and components along the squashed SE manifold \cite{Cassani:2010uw,Liu:2010sa,Gauntlett:2010vu,Skenderis:2010vz}. For the field strengths we define\footnote{If necessary, we will understand that the terms multiplying $\tD_{p>5}$ or $\tD_{<0}$ are equal to zero.}
\be\label{eq.decomp}
 H_{3} = \sum_{n=0}^3 H^{(3)}_{n} \w \tD_{3-n} , \quad F_{p} = \sum_{n=0}^p F^{(p)}_{n} \w \tD_{p-n} \ , \quad  \cF = \sum_{n=0}^2 \cF_{n} \w \tD_{2-n} \ .
\ee
A similar decomposition holds for all the potentials but $\AR$, since this gauge field is originated in the metric, where the SE structure is already explicit.


\section{Five-dimensional fields and action}\label{sec.action}

The decomposition (\ref{eq.decomp}) leads to the following five-dimensional fields.
The ten-dimensional RR one-form leads in five dimensions to
\bea
F^{(1)}_{0} &=&\customsign Q_f \ , \\ 
F^{(1)}_{1} &=& d C^{(0)}_{0} - \customsign Q_f \AR\ .
\eea
The ten-dimensional RR three-form gives
\bea\label{eq.F3defs}
F^{(3)}_{0}&=&F^{(1)}_{0} \cF_{0}\ , \\
F^{(3)}_{1} & = & dC^{(2)}_{0} - 2\sqrt{2} C^{(2)}_{1} - C^{(0)}_{0} H^{(3)}_{1} -\customsign Q_f \cF_{0} \AR \ ,   \\
F^{(3)}_{2} & = & dC^{(2)}_{1} - C^{(0)}_{0} H^{(3)}_{2}  +\customsign Q_f \left( \cF_{2} + \cF_{1} \w \AR \right) \ ,  \\
F^{(3)}_{3} & = & dC^{(2)}_{2} - C^{(0)}_{0} H^{(3)}_{3} - C^{(2)}_{1} \w d\AR -\customsign Q_f \cF_{2} \w \AR \ .
\eea
The self-dual ten-dimensional RR five-form provides the following five-dimensional fields
\bea
F^{(5)}_{0} & = & F^{(5)}_{fl} + \frac{F^{(1)}_{0}}{2} {\cF_{0}}^2 \ , \\
F^{(5)}_{1} & = & dC^{(4)}_{0} - 2\sqrt{2} C^{(4)}_{1} - \frac{1}{2} C^{(2)}_{0} H^{(3)}_{1} + \frac{1}{2} B^{(2)}_{0} \left(dC^{(2)}_{0} -2\sqrt{2} C^{(2)}_{1} \right) - F^{(5)}_{0} \AR \ , \\
F^{(5)}_{2} & = & dC^{(4)}_{1}  - \frac{1}{2} C^{(2)}_{0} H^{(3)}_{2} + \frac{1}{2} C^{(2)}_{1} \wedge H^{(3)}_{1}  + \frac{1}{2} B^{(2)}_{0} dC^{(2)}_{1}  \\
&& - \frac{1}{2} B^{(2)}_{1} \wedge \left( dC^{(2)}_{0}-2\sqrt{2} C^{(2)}_{1} \right) + \customsign Q_f \cF_{0}\left( \cF_{2} + \cF_{1}\w \AR \right) \ . \nonumber
\eea
The NSNS three-form decomposes to
\bea
H^{(3)}_{1} & =&  dB^{(2)}_{0} - 2\sqrt{2} B^{(2)}_{1}  \ , \label{eq.H3defs} \\
H^{(3)}_{2} & =&  dB^{(2)}_{1}  \ , \\
H^{(3)}_{3} & =&  dB^{(2)}_{2} -B^{(2)}_{1} \w d\AR \ .
\eea
Finally, the world-volume gauge field reduces to
\bea
\cF_{0} & = & 2\sqrt{2} \cA_{0} + B^{(2)}_{0}  \ , \label{eq.calFdefs} \\
\cF_{1} & = & d\cA_{0} + B^{(2)}_{1}  \ , \\
\cF_{2} & =&  d\cA_{1} +\cA_{0}d\AR + B^{(2)}_{2}  \ ,
\eea
where we have reabsorbed a factor of $2\pi\alpha' $ into the definition of $\cA$.

Thus, the ten-dimensional axion $C_{0}$ obviously reduces to a scalar $C^{(0)}_{0}$ in five dimensions.
The two-form RR potential $C_{2}$ is decomposed into a two-form $C^{(2)}_{2}$, a vector $C^{(2)}_{1}$ and a scalar $C^{(2)}_{0}$ with a St\"uckelberg coupling to the vector. For the NSNS two-form the decomposition is slightly different (apart from the two-form $B^{(2)}_{2}$) since there are two field strengths in which the vector $B^{(2)}_{1}$ couples to the exterior derivative of two different scalars: $B^{(2)}_{0}$ and $\cA_{0}$. In general we have the relation
\be
d \cF_{0}  =  H^{(3)}_{1} + 2\sqrt{2} \cF_{1}\ . \label{eq.F0F1H1}
\ee
The four-form $C_{4}$ reduces, due to the self-duality of $F_5$, just to a vector $C^{(4)}_{1}$ and a scalar $C^{(4)}_{0}$ with a St\"uckelberg coupling to the vector.
Finally, the vector $\cA$ gives, in five dimension, the vector $\cA_{1}$ and the scalar $\cA_{0}$.

We present in table \ref{tab.fields} a summary of the counting of fields in five dimensions, including the fields $g$, $f$, $w$ and $A_{1}$, coming from the ten-dimensional metric, and the dilaton $\Phi$.
\begin{table}[htbp]
  \centering
  \begin{tabular}{|c|c|c|c|c|}
    \hline
    Original 10d field & 5d two-forms & 5d vectors & 5d scalars & 5d metric \\ 
    \hline
    $C_{4}$ & $$ &  $C^{(4)}_{1}$ & $C^{(4)}_{0}$ & $$ \\
    $C_{2}$ & $C^{(2)}_{2}$ & $C^{(2)}_{1}$ & $C^{(2)}_{0}$ & $$ \\ 
    $C_{0}$ & $$ & $$ & $C^{(0)}_{0}$ & $$ \\ 
    $B_{2}$ & $B^{(2)}_{2}$ &$B^{(2)}_{1}$ & $B^{(2)}_{0}$ & $$ \\
    $\cA$ & $$ & $\cA_{1}$ & $\cA_{0}$ & $$ \\
    $G$ & $$ & $\AR$ & $f,w$ & $g$ \\
    $\Phi$ & $$ & $$ & $\Phi$ & $$ \\
    \hline
  \end{tabular}
  \caption{\label{tab.fields}Table of five-dimensional fields originating from the ten-dimensional ones. As discussed in the text, the scalars $C^{(2)}_{0}$, $C^{(4)}_{0}$ and $B^{(2)}_{0}$ are St\"uckelberg fields that might be set to zero, fixing the gauge and implying that the vectors $C^{(2)}_{1}$, $C^{(4)}_{1}$ and $B^{(2)}_{1}$ are massive, respectively. The vector $\cA_{1}$ corresponds to a pure gauge transformation of $B^{(2)}_{2}$ and can be gauged-fixed away.}
  \label{tab:label}
\end{table}

\vspace{0.2cm}
In order to write the action in a (almost) compact form, we also introduce the short-hand notation
\be
L_5 \equiv B_2 \w dC_2 - C_2 \w dB_2 + \cF^2 \w F_1^{D7} \ .
\ee
Decomposing in the basis given by the $\tD_{n}$ forms we obtain the components
\bea\label{L5bis}
L^{(5)}_{0} & = &  \customsign Q_f {\cF_{0}}^2 \ , \\
L^{(5)}_{1} & = & - C^{(2)}_{0}  \left( dB^{(2)}_{0}-2\sqrt{2}B^{(2)}_{1} \right)+  B^{(2)}_{0}  \left( dC^{(2)}_{0}-2\sqrt{2}C^{(2)}_{1} \right)   - \customsign  Q_f {\cF_{0}^2} \AR  \ , \\
L^{(5)}_{2} & = & - C^{(2)}_{0} dB^{(2)}_{1} + C^{(2)}_{1} \w \left( dB^{(2)}_{0}-2\sqrt{2}B^{(2)}_{1} \right)+  B^{(2)}_{0} dC^{(2)}_{1} \\
&&  -  B^{(2)}_{1} \w \left( dC^{(2)}_{0}-2\sqrt{2}C^{(2)}_{1} \right) + \customsign 2  Q_f \cF_{0} \left( \cF_{2} + \cF_{1}\w \AR \right) \ , \nonumber \\
L^{(5)}_{3} & = & - C^{(2)}_{0}  \left( dB^{(2)}_{2} - B^{(2)}_{1} \w d\AR \right) - C^{(2)}_{2} \w \left( dB^{(2)}_{0}-2\sqrt{2}B^{(2)}_{1} \right)\\
&& +  B^{(2)}_{0}  \left( dC^{(2)}_{2} - C^{(2)}_{1} \w d\AR \right)  +  B^{(2)}_{2} \w \left( dC^{(2)}_{0}-2\sqrt{2}C^{(2)}_{1} \right) - \customsign 2  Q_f \cF_{0} \cF_{2} \w \AR\ , \nonumber \\
L^{(5)}_{4} & = &  C^{(2)}_{1} \w  \left( dB^{(2)}_{2} - B^{(2)}_{1} \w d\AR \right) - C^{(2)}_{2} \w dB^{(2)}_{1} -  B^{(2)}_{1} \w \left( dC^{(2)}_{2} - C^{(2)}_{1} \w d\AR \right)\\ 
&& +  B^{(2)}_{2} \w dC^{(2)}_{1} +  \customsign Q_f \cF_{2} \w  \left( \cF_{2} + 2 \cF_{1}\w \AR \right)\ , \nonumber \\
L^{(5)}_{5} & = &  -C^{(2)}_{2} \w  \left( dB^{(2)}_{2} - B^{(2)}_{1} \w d\AR \right) + B^{(2)}_{2} \w  \left( dC^{(2)}_{2} - C^{(2)}_{1} \w d\AR \right) \label{L5bisend} \\
&& - \customsign Q_f \cF_{2} \w \cF_{2} \w \AR \ . \nonumber
\eea
Note that the $L^{(5)}_{i}$ are not independent fields but just a convenient notation.

Armed with these decompositions, we can reduce in a straightforward way the action (\ref{eq.IIBaction})-(\ref{eq.sources}) (see section \ref{action} and appendices \ref{app.topological}, \ref{app.DBI} for further details), obtaining the result in the next subsection.

\subsection{The action}\label{action}

We write here the result of applying the reduction ansatz to the ten-dimensional action given in \eqref{eq.newaction}. In principle this does not necessarily guarantee that the equations of motion derived from the variational principle in the reduced action coincide with the reduction of the ten-dimensional equations of motion. We have checked explicitly that this is the case and, for clarity, we present the action here before the reduced equations of motion.

To perform the reduction we need to express the ten-dimensional Hodge star operation, $\sten$, in terms of the five-dimensional one, $\sfive$, and the Hodge star in the SE manifold $*$. Taking into account the squashing and the conformal factor in front of the metric \eqref{eq.ansatzmetric2} we have
\bea\label{eq.Hodge10}
\sten \left( G_{p}\w \Delta_{n} \right) & = & (-1)^{n(5-p)} e^{\frac{25-10p}{3}f} e^{(2n-5)f} e^{[5((-1)^n-1)+2n]w} \left(\sfive G_{{p}} \right) \w \Delta_{5-n} \\
& = &(-1)^{n(5-p)}  \Gamma_{p,n} \left(\sfive G_{{p}} \right) \w \Delta_{5-n}  \ , \nonumber
\eea
where we have defined
\be\label{eq.Gammapn}
\Gamma_{p,n} = \exp\left[\frac{2}{3}(5(1-p)+3n)f+[5((-1)^n-1)+2n]w \right] \ ,
\ee
which in particular satisfies the relation
$
\Gamma_{n,5-n}\Gamma_{5-n,n}=1 
$.
Let us explain how the different factors of the scalars $f$ and $w$ arise. 
In the first line of \eqref{eq.Hodge10} the first exponent of $f$ is due to the conformal factor $\exp (10f/3)$ in front of the five-dimensional metric \eqref{eq.ansatzmetric2}.
The second exponent of $f$ takes into account the warping factor felt by the $\Delta_{5-n}$ on the r.h.s. (which contributes with an $(n-5)f$ factor) and the warping factor felt by the $\Delta_{n}$ on the l.h.s. (an additional $nf$ contribution). 
Finally, the exponent with the $w$ scalar receives the same $2n-5$ contribution as the term with $f$, and an additional $\pm 5w$ depending on whether $n$ is even (positive sign) or odd (negative sign). This is due to the fact that $\Delta_{2n+1}=\Delta_{2n}\w \Delta_{1}$, so the squashing has an effect in the presence of odd-degree $\Delta$'s.

Now we calculate the ten-dimensional kinetic terms, $G_{p}\w \sten G_{p}$, in the action. Omitting for the moment the dilaton and constant factors, we use \eqref{eq.Hodge10} to obtain
\be
\int G_{p} \w \sten G_{p} = V(X_{SE})  \sum_{n=0}^5 \int \Gamma_{n,p-n} G^{(p)}_{n} \w \sfive G^{(p)}_{n}  \ ,
\ee
where in the r.h.s. we have a five-dimensional integral, since we have performed the integral over the SE manifold, $\int_{SE} \Delta_{5} =  V(X_{SE})$. 

To reduce the Ricci scalar we use the orthogonal frame
\bea
\bar e^\alpha & = &e^{\frac{5}{3}f} e^\alpha\ ,  \phantom{e^{-(f+w)}e^i e^i} ( \alpha=0,\cdots,4 ) \label{eq.frame5d}\\
\bar e^i & = &e^{-(f+w)} e^i\ , \phantom{e^{\frac{5}{3}f}e^\alpha e^\alpha}  ( i=5,\cdots, 8 ) \\
\bar e^9 & = &e^{4w-f} (\eta+\AR) \ . \label{eq.frame9}
\eea
Notice that $\sqrt{2} \tD_{2} = J_{KE} = e^5 \w e^6 + e^7 \w e^8 = e^{2(f+w)} \left( \bar e^5 \w \bar e^6 + \bar e^7 \w \bar e^8 \right)$. The components of the Ricci tensor in ten dimensions can be read for example in \cite{Maldacena:2008wh} and are given by
\bea\label{eq.Riccibegin}
R^{(10)}_{\alpha\beta} & = & e^{-\frac{10}{3}f} \Bigg[ R^{(5)}_{\alpha\beta} - \frac{40}{3}\partial_\alpha f\partial_\beta f - 20 \partial_\alpha w \partial_\beta w - \frac{1}{2} e^{-\frac{16}{3}f+8w} (d\AR)_{\alpha\gamma} (d\AR)_\beta{^{\gamma}}  \\
&&\qquad \quad - \frac{5}{3} \eta_{\alpha\beta} \Box_5 f \Bigg] \ , \nonumber \\
R_{ij}^{(10)} & = & \delta_{ij} \left[6 e^{2f+2w}-2e^{2f+12w}+e^{-\frac{10}{3}f}\Box_5(f+w) \right] \ , \\
R_{99}^{(10)} & = & 4e^{2f+12w} - e^{-\frac{10}{3}f}\Box_5(4w-f) + \frac{1}{4} e^{-\frac{26}{3}f +8w} (d\AR)_{\alpha\beta}(d\AR)^{\alpha\beta}  \ , \\
R_{\alpha i}^{(10)} & = & R_{ i 9}^{(10)}  = 0 \ , \\
R_{\alpha 9}^{(10)} &  = & -\frac{1}{2} e^{-\frac{2}{3}f-4w} \nabla^\beta  \left(  e^{-\frac{16}{3}f+8w} (d\AR)_{\beta\alpha}\right) \ , \label{eq.Ricciend}
\eea
and since
$
\sten 1 = e^{\frac{10}{3}f} \left( \sfive 1\right) \w \Delta_{5}
$
 from \eqref{eq.Hodge10},
we are led to
\bea\label{eq.Riccis}
R^{(10)}\sten 1 &=& \Big[ R^{(5)} \sfive 1- \frac{40}{3} df \w \sfive df - 20 dw \w \sfive dw - \frac{1}{2} e^{-\frac{16}{3}f+8w} d\AR \w \sfive d\AR \\
&& \quad- 4 e^{\frac{16}{3}f+2w} \left(e^{10w} -6 \right)\sfive  1\Big] \w \Delta_{5} \ . \nonumber
\eea

Collecting the reductions of all the terms present in the action we obtain finally
\be\label{eq.reducedaction}
S_{5d}  =  \frac{1}{2\kappa_5^2} \int \left[  R \sfive1 -\sum \frac{k_i}{2}  e^{a_i \Phi+b_i f+c_i w} F_{n_i} \w \sfive F_{n_i}  - V \sfive1  \right] +S_{top} + S_{DBI,5d} \ ,
\ee
with $\kappa_5^2 = \kappa_{10}^2 / V{(X_{SE})}$.
In equation \eqref{eq.reducedaction}, the index $i$ runs over all matter fields present in the reduction, whose coefficients are given in table \ref{tab.actioncoeffs}.
\begin{table}[tb]
\begin{center}
\subtable{
\begin{tabular}{c|ccccc}
$F_{n_i}$ & $k_i$ & $a_i$ & $b_i$ & $c_i$ & $n_i$ \\
\hline
$df$ & $80/3$ & $0$ & $0$ & $0$ & $1$ \\ 
$dw$ & $40$ & $0$ & $0$ & $0$ & $1$  \\ 
$d\Phi$ & $1$ & $0$ & $0$ & $0$ & $1$  \\
\hline 
$d\AR$ & $1$ & $0$ & $-16/3$ & $8$ & $2$  \\ 
\hline 
$F^{(1)}_{1}$ & $1$ & $2$ & $0$ & $0$ & $1$  \\ 
\hline
$F^{(5)}_{1}$ & $1$ & $0$ & $8$ & $8$ & $1$  \\ 
$F^{(5)}_{2}$ & $1$ & $0$ & $8/3$ & $-4$ & $2$  
\end{tabular}
}\hspace{1cm}
\subtable{
\begin{tabular}{c|ccccc}
$F_{n_i}$ & $k_i$ & $a_i$ & $b_i$ & $c_i$ & $n_i$ \\
\hline
$F^{(3)}_{1}$ & $1$ & $1$ & $4$ & $4$ & $1$ \\ 
$F^{(3)}_{2}$ & $1$ & $1$ & $-4/3$ & $-8$ & $2$ \\ 
$F^{(3)}_{3}$ & $1$ & $1$ & $-20/3$ & $0$ & $3$ \\ 
\hline
$H^{(3)}_{1}$ & $1$ & $-1$ & $4$ & $4$ & $1$ \\ 
$H^{(3)}_{2}$ & $1$ & $-1$ & $-4/3$ & $-8$ & $2$  \\ 
$H^{(3)}_{3}$ & $1$ & $-1$ & $-20/3$ & $0$ & $3$ 
\end{tabular}
}
\caption{Coefficients associated to the forms that enter in the reduced five-dimensional action \eqref{eq.reducedaction}. Notice that these correspond to kinetic terms, not the potentials. \label{tab.actioncoeffs}}
\end{center}
\end{table}
Comparing with the reduction of the form products appearing in \eqref{eq.newaction}, and the relation between the ten-dimensional and five-dimensional Ricci scalar \eqref{eq.Riccis}, we find that the potential is given by
\be
V = 4 e^{\frac{16}{3}f+2w} \left( e^{10w}-6 \right) + \frac{1}{2} e^{\frac{40}{3}f} \left( F^{(5)}_{0} \right)^2  +  \frac{1}{2} e^{\Phi+\frac{28}{3}f-4w} \left( F^{(3)}_{0} \right)^2  + \frac{1}{2} e^{2\Phi + \frac{16}{3}f - 8 w} \left( F^{(1)}_{0} \right)^2 \ .
\ee
It is worth noting that for $Q_f \sim {\cal O}(1)$ the potential does not admit any standard $AdS$ vacuum.
The topological part of the action, $S_{top}$, is given by (see appendix \ref{app.topological} for details on the derivation)
\bea\label{eq.topological}
S_{top} & = & \frac{1}{2\kappa_{5}^2} \int  \frac{1}{2}   C^{(4)}_{1} \w d \AR \w dC^{(4)}_{1}    +\customsign^2 \frac{Q_f^2}{8} {\cF_{0}}^2 {\cF_{2}}^2 \w \AR -  \frac{1}{2}  F^{(5)}_{fl}  L^{(5)}_{5}  \\
&& \qquad \quad  +  \frac{1}{4}  \left(F^{(5)}_{1}+dC^{(4)}_{0} - 2\sqrt{2} C^{(4)}_{1} -F^{(5)}_{fl}\AR  + \customsign\frac{Q_f}{2} {\cF_{0}}^2 \AR \right)\w L^{(5)}_{4}  \nonumber\\
&& \qquad \quad -  \frac{1}{4}  \left(F^{(5)}_{2}+dC^{(4)}_{1} -\customsign Q_f \cF_{0} \left(\cF_{2} + \cF_{1} \w \AR \right)  \right)\w L^{(5)}_{3} \nonumber\\
&& \qquad \quad  + \customsign \frac{Q_f}{4} \cF_{0} \cF_{2} \w \AR \w L^{(5)}_{2}  + \customsign \frac{Q_f}{8} \cF_{2} \w \left( \cF_{2} + 2 \cF_{1} \w \AR \right) \w L^{(5)}_{1}   \nonumber\\
&& \qquad\quad + \customsign \frac{Q_f}{4} {\cF_{0}}^2 \left( F^{(1)}_{1} \w \cF_{2} \w \left( \cF_{2} + 2 \cF_{1} \w \AR \right)  + F^{(1)}_{0} {\cF_{2}}^2 \w \AR \right)   \nonumber \\
&& \qquad \quad- \customsign \frac{Q_f}{2} \cF_{0} \Bigg(  \cF_{0} F^{(3)}_{3} \w \left( \cF_{2} + \cF_{1} \w \AR \right)  +  \cF_{0} F^{(3)}_{2} \w  \cF_{2} \w \AR  \nonumber \\
&& \qquad \quad \qquad \qquad \quad + F^{(3)}_{1} \w \cF_{2} \w \left( \cF_{2} + 2 \cF_{1} \w \AR \right)  + F^{(3)}_{0} {\cF_{2}}^2 \w \AR \Bigg)  
 \ ,\nonumber
\eea
whereas the five-dimensional reduction of the DBI term reads (details on how to obtain this expression can be found in  appendix \ref{app.DBI})
\be
S_{DBI,5d} = -\frac{4Q_f}{2\kappa_{5}^2} \int e^{\Phi+\frac{16}{3}f+2w}\sqrt{\det(g^{-1}\cdot Z)} \sqrt{ 1+\frac{e^{4f+4w-\Phi}}{2}   {\cF_{0}}^2 }\, \sfive 1 \ ,
\ee
with $Z$ a matrix that can be written formally as
\be\label{eq.Zdef}
Z=g+ e^{-\frac{\Phi}{2}-\frac{10}{3}f} \left( \cF_{2} + \cF_{1} \w \AR \right)  + e^{-\Phi-\frac{4}{3}f-8w}\cF_{1}\otimes \cF_{1}   \ ,
\ee
$g$ being the five-dimensional metric.

\section{Five-dimensional equations of motion}\label{sec.eqs}

The equations of motion in five dimensions can be derived either from the action (\ref{eq.reducedaction}) or from the reduction of the ten-dimensional equations \eqref{eqforF1}-\eqref{eqforR}. In both cases one arrives to the same result.

For the differential forms the simplest way to obtain the reduced equations is to directly reduce the equations of motion. 
From the $F_{1}$ field-strength equation (\ref{eqforF1}) we get
\bea
d\left[ e^{2\Phi} \sfive  F^{(1)}_{1} \right] & = & - e^{ - \frac{20}{3} f + \Phi } \sfive F^{(3)}_{3} \w H^{(3)}_{3} - e^{  -\frac{4}{3} f - 8 w + \Phi } \sfive F^{(3)}_{2} \w H^{(3)}_{2}\label{eq.F11} \\
&& - e^{ 4( f + w ) + \Phi } \sfive F^{(3)}_{1} \w H^{(3)}_{1} +\sqrt{2} \,\customsign\,  Q_f \cF_{0} \cF_{2} \w \cF_{2} \w \cF_{1}  \ , \nonumber
\eea
and there is no equation of motion for $F^{(1)}_{0}$, which is a flux due to the presence of D7 (anti-) branes.

Equation (\ref{eqforF3}) for $F_{3}$ results in three different five-dimensional equations, of which only two are independent
\bea\label{eq.F33}
d\left[ e^{\Phi-\frac{20}{3}f} \sfive F^{(3)}_{3} \right] & = & F^{(5)}_{1} \w H^{(3)}_{2} \!-  F^{(5)}_{2} \w H^{(3)}_{1} \!-  F^{(5)}_{0} \w H^{(3)}_{3}  \! - 2\sqrt{2} \customsign Q_f \cF_{0}\, \cF_{2} \w \cF_{1} \ ,\\
d\left[ e^{\Phi-\frac{4}{3}f-8w} \sfive F^{(3)}_{2} \right] & = & 2 \sqrt{2} e^{\Phi+4(f+w)} \sfive F^{(3)}_{1} + e^{\Phi-\frac{20}{3}f} \sfive F^{(3)}_{3} \w d\AR   -  F^{(5)}_{1} \w H^{(3)}_{3}  \label{eq.F32} \\
&& +e^{\frac{8}{3}f-4w} \sfive F^{(5)}_{2} \w H^{(3)}_{1} + \sqrt{2} \,  \customsign \, Q_f \cF_{0} \cF_{2} \w \cF_{2} \nonumber
 \ . 
\eea
From \eqref{eq.F32} we can obtain the equation of motion for $C^{(2)}_{0}$ by acting with the exterior derivative. This is a consequence of the St\"uckelberg coupling between the five-dimensional scalar and vector associated to the reduction of $C_{2}$: $C^{(2)}_{0}$ can be seen as the longitudinal component of the vector $C^{(2)}_{1}$.

Equation (\ref{eqforF5}) for $F_{5}$, after self-duality has been considered, gives just one independent equation of motion for the reduced vector potential $C^{(4)}_{1}$
\bea
d\left[ e^{\frac{8}{3}f-4w} \sfive F^{(5)}_{2}  \right]  & = &  2\sqrt{2} e^{8f+8w} \sfive F^{(5)}_{1} + F^{(5)}_{2} \w d\AR-H^{(3)}_{1} \w F^{(3)}_{3} -H^{(3)}_{3} \w F^{(3)}_{1}    \label{eq.F52} \\
&&  - \sqrt{2}\, \customsign Q_f \cF_{2} \w \cF_{2}  
 \  . \nonumber
\eea
Once again, acting with the exterior derivative produces the equation of motion for $C^{(4)}_{0}$ due to the St\"uckelberg coupling between the scalar and vector in the reduction of $C_{4}$.
The forms $C^{(4)}_{i}$ with $i=2,3,4$ are given in terms of $C^{(4)}_{0}$, $C^{(4)}_{1}$ by the self-duality of $F_{5}$.

In the NSNS sector the presence of the DBI term in the action becomes manifest, since the two-form potential $B_{2}$ couples to the gauge field in the world-volume of the D7-branes via the gauge-invariant combination $\cF$. The effect of such term in the Lagrangian is denoted in the reduced equations  by the presence of $n$-forms, $\Theta_{n}$, whose precise definition is relegated to appendix \ref{app.NSNS}. We just want to note that the $\Theta_{n}$ are proportional to $Q_f$ but insensitive to the factor $\customsign$, since their origin is tracked to the DBI action, not the WZ term. The equations of motion are
\bea
d\left[ e^{-\Phi-\frac{20}{3}f} \sfive H^{(3)}_{3} \right] & = &  e^{\Phi-\frac{20}{3}f} F^{(1)}_{1} \w \sfive F^{(3)}_{3} + e^{\Phi-\frac{4}{3}f-8w} F^{(1)}_{0} \sfive F^{(3)}_{2}  - F^{(5)}_{1} \w F^{(3)}_{2}   \label{eq.H33} \\
&&+  e^{\frac{8}{3}f-4w} \sfive F^{(5)}_{2} \w F^{(3)}_{0}+ F^{(5)}_{2} \w F^{(3)}_{1} +  F^{(5)}_{0}  F^{(3)}_{3} - \Theta_{3} \ ,\nonumber \\
d\left[ e^{-\Phi-\frac{4}{3}f-8w} \sfive H^{(3)}_{2} \right] & = & 2 \sqrt{2} e^{-\Phi+4(f+w)} \sfive H^{(3)}_{1}  + e^{-\Phi-\frac{20}{3}f} \sfive H^{(3)}_{3} \w d\AR +  F^{(5)}_{1} \w F^{(3)}_{3}    \label{eq.H32} \\
&& +  e^{\Phi-\frac{4}{3}f-8w}F^{(1)}_{1}\w \sfive F^{(3)}_{2}  - e^{\frac{8}{3}f-4w} \sfive F^{(5)}_{2} \w F^{(3)}_{1}+ \Theta_{4}\ , \nonumber \\
d\left[ e^{-\Phi+4f+4w} \sfive H^{(3)}_{1} \right] & = & e^{\Phi +4f+4w}  F^{(1)}_{1} \w \sfive F^{(3)}_{1}  +  e^{\Phi+\frac{28}{3}f-4w}F^{(1)}_{0}\w \sfive F^{(3)}_{0}  +  F^{(5)}_{2} \w F^{(3)}_{3}  \label{eq.H31} \\
&& +  e^{\frac{40}{3}f}\sfive F^{(5)}_{0} \w F^{(3)}_{0}  +  e^{8f+8w}\sfive F^{(5)}_{1} \w F^{(3)}_{1}+  e^{\frac{8}{3}f-4w}\sfive F^{(5)}_{2} \w F^{(3)}_{2} \nonumber \\
&&  - \Theta_{5}\ , \nonumber
\eea
with, in coordinate basis and with indices in round (square) parenthesis indicating symmetrization (antisymmetrization),
\bea\label{theta3}
\Theta_{3} & = & - \frac{4Q_f }{2!\, 3!} e^{\frac{\Phi}{2} + 2f + 2w} \sqrt{|Z|} \sqrt{ 1 + \frac{e^{-\Phi+4f+4w}}{2} {\cF_{0}}^2 } \,(Z^{-1})^{ba} \epsilon_{abc_1c_2c_3} dx^{c_1}\w dx^{c_2} \w dx^{c_3} \ , \\\label{theta4}
\Theta_{4} & = &  - \frac{4Q_f }{4!} e^{\frac{\Phi}{2} + 2f + 2w} \sqrt{|Z|} \sqrt{ 1 + \frac{e^{-\Phi+4f+4w}}{2} {\cF_{0}}^2 } \\ 
&& \,\,  \times \left[  (Z^{-1})^{[ba]}  {\AR}_b + 2 e^{-\frac{\Phi}{2}+2f-8w} (Z^{-1})^{(ba)}  {\cF_{1}}_b  \right]  \epsilon_{ac_1c_2c_3c_4} dx^{c_1}\w dx^{c_2} \w dx^{c_3} \w dx^{c_4}  \ , \nonumber\\\label{theta5}
\Theta_{5} & = & - 2 Q_f  e^{ \frac{28}{3} f +6w} \sqrt{|g^{-1}\cdot Z|}  \left[ 1 + \frac{e^{-\Phi+4f+4w}}{2} {\cF_{0}}^2 \right]^{-1/2} \sfive \cF_{0} \ .
\eea
Contrary to the case of the RR two-form potential, equation \eqref{eq.H31} is not implied by \eqref{eq.H32} upon acting with the exterior derivative. The ultimate reason is that even if the dimensional reduction gives the St\"uckelberg coupling $H^{(3)}_{1}=dB^{(2)}_{0}-2\sqrt{2} B^{(2)}_{1}$, the vector field $B^{(2)}_{1}$ is also coupled to a different scalar through the relation $\cF_{1}= d\cA_{0}+B^{(2)}_{1}$. Indeed, this is the origin of the relation \eqref{eq.F0F1H1}. A specific calculation showing the independence of equations \eqref{eq.H33}-\eqref{eq.H31} (and the relation to the equations of motion for $\cA_{0},\cA_{1}$) is found in  appendix \ref{app.NSNS}.

Thus, reducing the ten-dimensional equation of motion for the NSNS potential we obtain three equations, \eqref{eq.H33}-\eqref{eq.H31}, which are all independent and must be used to find $B^{(2)}_{0,1,2}$ and $\cA_{0,1}$. As mentioned, both $B^{(2)}_{0}$ and $\cA_{0}$ couple in a St\"uckelberg fashion to the vector $B^{(2)}_{1}$, so one of the two scalars corresponds to a gauge degree of freedom, whereas the second scalar is a physical degree of freedom.
Moreover, $B^{(2)}_{2}$ enters in the action only via its external derivative or added to $d\cA_{1}$, which can be interpreted as a gauge transformation of $B^{(2)}_{2}$.
The implication is that considering a non-trivial $\cA_{1}$ can be traded for a gauge fixing in $B^{(2)}_{2}$.
Therefore, equations \eqref{eq.H33}-\eqref{eq.H31} suffice to describe the ten-dimensional NSNS potential and D7-worldvolume gauge degrees of freedom in the effective five-dimensional theory.

The last ten-dimensional matter field to consider is the dilaton. Once again, the existence of the DBI term in the total action is reflected in the equation of motion, which in this case is derived easily from the five-dimensional action \eqref{eq.reducedaction} as
\be
d\sfive d\Phi = \frac{1}{2} \sum k_i a_i e^{a_i \Phi+b_i f+c_i w} F_{n_i} \w \sfive F_{n_i} + \frac{\partial V}{\partial \Phi} \sfive 1 - 2\kappa_5^2 \frac{\delta S_{DBI,5d}}{\delta \Phi}  \ ,
\ee
which gives the same result as the ten-dimensional equation of motion (\ref{eqforP}) expressed in our five-dimensional decomposition
\bea\label{eq.reddil}
d \sfive d\Phi & = & -\frac{1}{2} e^{-\Phi} \left(  e^{-\frac{20}{3}f} H^{(3)}_{3} \w \sfive H^{(3)}_{3} + e^{-\frac{4}{3}f-8w} H^{(3)}_{2} \w \sfive H^{(3)}_{2}+ e^{4f+4w} H^{(3)}_{1} \w \sfive H^{(3)}_{1} \right) \\
&& + \frac{1}{2} e^{\Phi} \left(  e^{-\frac{20}{3}f} F^{(3)}_{3} \w \sfive F^{(3)}_{3} + e^{-\frac{4}{3}f-8w} F^{(3)}_{2} \w \sfive F^{(3)}_{2}+ e^{4f+4w} F^{(3)}_{1} \w \sfive F^{(3)}_{1} \right. \nonumber \\ 
&& \left. + e^{\frac{28}{3}f-4w} F^{(3)}_{0} \w \sfive F^{(3)}_{0} \right) + e^{2\Phi} F^{(1)}_{1} \w \sfive F^{(1)}_{1} +e^{2\Phi+\frac{16}{3}f-8w} F^{(1)}_{0} \w \sfive F^{(1)}_{0} \nonumber \\
&& -Q_f e^{\frac{28}{3}f+6 w} \frac{   \sqrt{\det(g^{-1}\cdot Z)}    {\cF_{0}}^2}{\sqrt{1+\frac{1}{2} e^{-\Phi+4f+4w} {\cF_{0}}^2}} \sfive 1 \nonumber \\
&& + 4Q_f e^{\Phi+\frac{16}{3}f +2w} \sqrt{\det (g^{-1}\cdot Z)}   \sqrt{ 1+\frac{{\cF_{0}}^2}{2}e^{-\Phi+4f+4w} } \left( 1+ \frac{1}{2} \text{tr} \left[ Z^{-1}\cdot \frac{\delta Z}{\delta \Phi} \right] \right)\sfive 1  \ .\nonumber
\eea

Similarly, we can find the equations of motion for the scalars $f$ and $w$ as
\bea\label{eq.feom}
\frac{80}{3}\, d\sfive df &  = &\frac{1}{2} \sum k_i b_i e^{a_i \Phi+b_i f+c_i w} F_{n_i} \w \sfive F_{n_i} + \frac{\partial V}{\partial f} \sfive 1 - 2\kappa_5^2 \frac{\delta   S_{DBI,5d} }{\delta f} \ , \\
40 \, d\sfive dw & = & \frac{1}{2} \sum k_i c_i e^{a_i \Phi+b_i f+c_i w} F_{n_i} \w \sfive F_{n_i} + \frac{\partial V}{\partial w} \sfive 1 - 2\kappa_5^2 \frac{\delta  S_{DBI,5d} }{\delta w} \ , \label{eq.weom}
\eea
which are given explicitly by
\bea\label{eqforfexpl}
40\, d \sfive df & = & 6 e^{8f+8w} F^{(5)}_{1}\w\sfive F^{(5)}_{1}+ 2 e^{\frac{8}{3}f-4w} F^{(5)}_{2}\w\sfive F^{(5)}_{2} - 4 e^{-\frac{16}{3}f+8w}d\AR\w\sfive d\AR  \\
&& + e^{\Phi} \left[3 e^{4f+4w} F^{(3)}_{1}\w\sfive F^{(3)}_{1} -  e^{-\frac{4}{3}f-8w} F^{(3)}_{2}\w\sfive F^{(3)}_{2} -5 e^{-\frac{20}{3}f} F^{(3)}_{3}\w\sfive F^{(3)}_{3} \right] \nonumber \\
&& + e^{-\Phi} \left[3 e^{4f+4w} H^{(3)}_{1}\w\sfive H^{(3)}_{1} -  e^{-\frac{4}{3}f-8w} H^{(3)}_{2}\w\sfive H^{(3)}_{2} -5 e^{-\frac{20}{3}f} H^{(3)}_{3}\w\sfive H^{(3)}_{3} \right] \nonumber \\
&& + \Bigg[ 32 e^{\frac{16}{3}f+2w}\left( e^{10w}-6\right) + 10 e^{\frac{40}{3}f}(F^{(5)}_{0})^2  + 7 e^{\Phi+\frac{28}{3}f-4w}(F^{(3)}_{0})^2  \nonumber\\
&& + 4 e^{2\Phi+\frac{16}{3}f -8w}(F^{(1)}_{0})^2 \Bigg] \sfive 1  +6Q_f e^{\frac{28}{3}f+6 w}  \frac{ \sqrt{\det(g^{-1}\cdot Z)}    {\cF_{0}}^2}{\sqrt{1+\frac{1}{2} e^{-\Phi+4f+4w} {\cF_{0}}^2}} \sfive 1 \nonumber  \\
&& +6 Q_f e^{\Phi+\frac{16}{3}f+2w} \sqrt{\det(g^{-1}\cdot Z)}  \sqrt{1+\frac{{\cF_{0}}^2}{2} e^{-\Phi+4f+4w} }  \left( \frac{16}{3}+\text{tr} \left[ Z^{-1}\cdot \frac{\delta Z}{\delta f} \right] \right) \sfive 1 \ ,\nonumber 
\eea
and
\bea\label{eqforwexpl}
40\, d \sfive dw & = & 4 e^{8f+8w} F^{(5)}_{1}\w\sfive F^{(5)}_{1} - 2 e^{\frac{8}{3}f-4w} F^{(5)}_{2}\w\sfive F^{(5)}_{2} + 4 e^{-\frac{16}{3}f+8w}d\AR\w\sfive d\AR  \\
&& + e^{\Phi} \left[2 e^{4f+4w} F^{(3)}_{1}\w\sfive F^{(3)}_{1} - 4 e^{-\frac{4}{3}f-8w} F^{(3)}_{2}\w\sfive F^{(3)}_{2} \right] \nonumber \\
&& + e^{-\Phi} \left[2 e^{4f+4w} H^{(3)}_{1}\w\sfive H^{(3)}_{1} - 4 e^{-\frac{4}{3}f-8w} H^{(3)}_{2}\w\sfive H^{(3)}_{2} \right] \nonumber \\
&& + \Bigg[ 48 e^{\frac{16}{3}f+2w}\left( e^{10w}-1 \right) -2 e^{\Phi+\frac{28}{3}f-4w}(F^{(3)}_{0})^2   - 4 e^{2\Phi+\frac{16}{3}f -8w}(F^{(1)}_{0})^2 \Bigg] \sfive 1 \nonumber\\
&& +4Q_f e^{\frac{28}{3}f+6 w} \frac{   \sqrt{\det(g^{-1}\cdot Z)}    {\cF_{0}}^2}{\sqrt{1+\frac{1}{2} e^{-\Phi+4f+4w} {\cF_{0}}^2}} \sfive 1 \nonumber  \\
&& +2 Q_f e^{\Phi+\frac{16}{3}f+2w} \sqrt{\det(g^{-1}\cdot Z)}  \sqrt{1+\frac{{\cF_{0}}^2}{2} e^{-\Phi+4f+4w} }  \left( 4 +\text{tr} \left[ Z^{-1}\cdot \frac{\delta Z}{\delta w} \right] \right) \sfive 1 \ .\nonumber 
\eea

The equation of motion we get for the $U(1)_R$ gauge field receives also a contribution from the DBI term, which is derived in appendix \ref{app.DBI}. The resulting equation is found easily from the effective five-dimensional action and reads
\bea\label{eq.U1Reom}
d\left[ e^{-\frac{16}{3}f+8w} \sfive d \AR \right] & = &  e^{2\Phi}F^{(1)}_{0} \w \sfive F^{(1)}_{1} +e^{\Phi+4f+4w} F^{(3)}_{0} \w \sfive F^{(3)}_{1}+e^{\Phi-\frac{20}{3}f  } F^{(3)}_{2} \w \sfive F^{(3)}_{3}    \\
&& + e^{-\Phi-\frac{20}{3}f } H^{(3)}_{2} \w \sfive H^{(3)}_{3} + e^{8f+8w} F^{(5)}_{0} \w \sfive F^{(5)}_{1}  + \frac{1}{2} F^{(5)}_{2} \w  F^{(5)}_{2} \nonumber\\
&& + d \left( \cA_{0} \Theta_3 \right) - \cF_{1}\w \Theta_3 \ ,\nonumber
\eea
where $d\Theta_3$ can be obtained from equation \eqref{eq.H33}. Acting with the exterior derivative in \eqref{eq.U1Reom}, as it stands, is only compatible with $d^2=0$ if $dQ_f=0$, as was assumed all over. Considering the case with $dQ_f\neq0$ would have required the presence of fields charged under $\AR$ in our reduction ansatz \cite{Cassani:2010uw,Gauntlett:2010vu}.

Finally, the five-dimensional Einstein equations read
\bea
R^{\alpha\beta} & = & \frac{1}{2} \sum k_i e^{a_i \Phi+b_i f+c_i w} \left[ \left( \iota^\alpha F_{n_i} \right) \lrcorner \left( \iota^\beta F_{n_i} \right) - \frac{n_i-1}{3} g^{\alpha\beta} F_{n_i} \lrcorner F_{n_i}  \right]  + \frac{V}{3}g^{\alpha\beta} \\
&& - 2 Q_f e^{\Phi+\frac{16}{3}f+2w}\sqrt{\det  g^{-1}\cdot Z} \sqrt{ 1+\frac{e^{4f+4w-\Phi}}{2}   {\cF_{0}}^2 } \left[ \left(Z^{-1} \right)^{(\alpha\beta)} - \frac{1}{3}g^{\alpha\beta}  \,\text{tr} \left[  Z^{-1}\cdot g \right] \right] \ . \nonumber
\eea

To check the consistency of the reduction one should obtain from the ten-dimensional equations of motion for the metric \eqref{eq.ansatzmetric2} the equations of motion for $f$, $w$, $\AR$ and the five-dimensional Einstein equations. The ten-dimensional Einstein equations expressed in Ricci form can be obtained using the explicit components given in \eqref{eq.Riccibegin}-\eqref{eq.Ricciend}.

Since the $R^{(10)}_{ij}$ components are proportional to $\delta_{ij}$, the Einstein equations in these directions give only one independent equation of motion. The $R^{(10)}_{99}$ component gives a second one, and considering two combinations of them we arrive to the equations of motion \eqref{eqforfexpl} and \eqref{eqforwexpl} for the scalars $f$ and $w$ respectively. The $R^{(10)}_{\alpha 9}$ equation gives equation \eqref{eq.U1Reom}, and $R_{i 9}^{(10)} = R_{\alpha i}^{(10)}  = 0$ are identically satisfied by evaluating the r.h.s. of the Einstein equations in ten dimensions. The five-dimensional Einstein equations are just the $R_{\alpha\beta}^{(10)}$ components of the ten-dimensional ones.

%
%

\section{Comments on the dual operators and sub-truncations}\label{sec.comments}

The reduction we have performed includes the R-uncharged, un-flavored spectrum derived in \cite{Cassani:2010uw,Gauntlett:2010vu} from the Type IIB theory on squashed Sasaki-Einstein manifolds.
We can refer to the analysis in that paper and in \cite{Ceresole:1999zs} for the discussion of the operators of the dual theories.

This setting is extended in our case to include a few operators from the flavor sector.
These are discussed, in the quenched ${\cal N}=4$ case, in \cite{Aharony:1998xz}, \cite{Kruczenski:2003be}.
On the standard $AdS$ vacuum the massless vector $\cA_{1}$ corresponds to the flavor current of conserved dimension $\Delta=3$.
The dual operator is of the form $\bar \psi_i^\alpha \gamma^{\mu}_{\alpha\beta} \psi^{i \beta} + i \bar q^a (D^{\mu} q_a) -i (\bar D^{\mu} \bar q^a) q_a $, where $q_a, \psi_i$ are the squarks and quarks in the fundamental (see e.g. \cite{Erdmenger:2007cm}).

From the five-dimensional action of the previous sections we calculate that the scalar $\cA_{0}$ has $m^2=-4$ (we are considering the $\customsign=-1$ brane case), corresponding to the $\Delta=2$ operator of the form $\bar q^a \sigma^3_{ab} q^b$, $\sigma^3$ being the Pauli matrix \cite{Aharony:1998xz,Kruczenski:2003be,Erdmenger:2007cm}. 
This operator sits in a vector multiplet together with the operator dual to the vector $\cA_{1}$, which has protected dimension three; thus, the operator dual to  $\cA_{0}$ has dimension two. 
This is an interesting mode, since it sits at the edge of the BF bound and its quasinormal mode at zero temperature, finite chemical potential in the probe approximation has a pole precisely on the real axis \cite{Ammon:2011hz}.

\vspace{0.2cm}
It is interesting to consider the subset of operators corresponding to the charged solution in \cite{Bigazzi:2011it} beyond the quenched limit.
The flavored solution at zero charge corresponds to the inclusion of vevs for a dimension eight operator of the form Tr$F^4$ dual to the gravity scalar $f$, a dimension six operator of the form Tr$({\cal W}_\alpha {\cal W}^\alpha)^2$ dual to the gravity scalar $w$, and the insertion of the marginally irrelevant flavor term in the Lagrangian dual to the gravity scalar $\Phi$ \cite{Benini:2006hh}.

The charged solution in \cite{Bigazzi:2011it} includes also the gravity fields $C^{(2)}_{2}, C^{(2)}_{1}, \cA_{1}$.
On the $AdS$ vacuum\footnote{We think about $AdS$ as a convenient vacuum for the analysis of the spectrum in the small $Q_f$ limit, even if the flavored theory at finite $Q_f$ does not strictly admit it as a standard vacuum (the dilaton has a run-away behavior).} and in the ``unquenched'' case the equations for their fluctuations are coupled to the one of $B^{(2)}_{2}$ and read\footnote{In the equation for $C^{(2)}_{1}$ we gauged to zero the St\"uckelberg field $C^{(2)}_{0}$.}
\bea\label{AdSfluc1}
d\left( \sfive  dC^{(2)}_{2} \right) & = & -Q_c dB^{(2)}_{2} \ , \\ 
\label{AdSfluc2}
d\left( \sfive  dB^{(2)}_{2} \right) & = & \customsign Q_f \sfive \left( dC^{(2)}_{1} + \customsign Q_f \cF_{2} \right) + Q_c dC_{2}+8Q_f\sfive \cF_{2} \ , \\ 
\label{AdSfluc3}
d \sfive \left( dC^{(2)}_{1} + \customsign Q_f \cF_{2} \right)  & = &  -8 \sfive C^{(2)}_{1} \ ,\\ 
\label{AdSfluc4}
8 d\sfive \cF_{2}  & = & - \customsign d \sfive \left( dC^{(2)}_{1} + \customsign Q_f \cF_{2} \right) \ ,
\eea
where
\bea
\cF_{2}=d\cA_{1}+B^{(2)}_{2}\ .
\eea
We have made explicit the term $F^{(5)}_{fl}\equiv Q_c=4R_{AdS}=4$ in units where the radius of $AdS$ is one  \cite{Benini:2006hh}.\footnote{After equation \eqref{eq.ansatzmetric2} we have assumed that the radius of the (squashed) SE manifold, $R_{SE}$, was set to one, and this seems to contradict our use of $R_{AdS}=1$ if $R_{AdS}\neq R_{SE}$. However, these two can be made to coincide by a constant shift in $f$.}
 
The relevant part of the action for these fluctuations is
\bea\label{acflucads}
S &\sim & \int -\frac12 dC^{(2)}_{2} \wedge \sfive dC^{(2)}_{2} - \frac12 \left(dC^{(2)}_{1} + \customsign Q_f \cF_{2} \right)\wedge \sfive \left( dC^{(2)}_{1}  + \customsign Q_f \cF_{2} \right) \\
&& \quad- 4 C^{(2)}_{1} \wedge \sfive C^{(2)}_{1}  -\frac12 dB^{(2)}_{2} \wedge \sfive dB^{(2)}_{2}  + Q_cdB^{(2)}_{2} \wedge C^{(2)}_{2}-4Q_f \cF_{2}\wedge\sfive\cF_{2}\ . \nonumber
\eea

In order to read the field content of the action (\ref{acflucads}) we have to disentangle the various modes.\footnote{We thank Alberto Zaffaroni for a very nice discussion about what follows.}
This can be achieved by standard procedures, see e.g. \cite{Gauntlett:2009zw}.
As a first step, we dualize $\cA_{1}$ by introducing the Lagrange multiplier $\tilde B_{2}$; after defining $d\cA_{1}\equiv \hat F_{2}$, we add to the action the term
\bea\label{lagrange1}
 -(8Q_f+\customsign^2 Q_f^2) \int \tilde B_{2}\wedge d \hat F_{2}\ .
\eea
The equation of motion for $\hat F_{2}$ from (\ref{acflucads}), (\ref{lagrange1}) gives
\bea
\hat F_{2}=-B^{(2)}_{2}-\sfive d\tilde B_{2} - \frac{\customsign Q_f}{8 Q_f+\customsign^2 Q_f^2} dC^{(2)}_{1}
\eea
and integrating $\hat F_{2}$ out leaves us with
\bea\label{acflucads1}
S  &  \sim & \int  -\frac12 dC^{(2)}_{2} \wedge \sfive dC^{(2)}_{2} -\frac12 dB^{(2)}_{2} \wedge \sfive dB^{(2)}_{2} -\frac12 (8Q_f +\customsign^2 Q_f^2) d\tilde B_{2}\wedge\sfive d\tilde B_{2}\\
&& \quad +dB^{(2)}_{2} \wedge \left[ Q_c C^{(2)}_{2} + (8Q_f +\customsign^2 Q_f^2)\tilde B_{2}\right]  -\frac{4}{8+\customsign^2 Q_f} dC^{(2)}_{1} \wedge \sfive dC^{(2)}_{1}-4C^{(2)}_{1} \wedge \sfive C^{(2)}_{1}\ . \nonumber
\eea

After canonical normalization, $C^{(2)}_{1}$ is recognized to be a massive vector with mass squared $m^2=8+\customsign^2 Q_f$, corresponding to a $\Delta=2+\sqrt{9+\customsign^2 Q_f}$ operator. At $Q_f=0$ the latter sits in the $\Delta=9/2$ supermultiplet of Tr$(\bar {\cal W}_{\dot\alpha} {\cal W}_\beta{\cal W}^\beta)+ \cdots$ \cite{Cassani:2010uw}, \cite{Ceresole:1999zs}. 
Thus, the operator is schematically of the form $\lambda \lambda F_{\dot\alpha\dot\beta}+c.c.+\bar \lambda_{\dot\alpha}F_{\dot\beta\beta}\lambda^{\beta}+ \cdots$ where $F_{\alpha\beta}$ is the self dual part of the gauge vector field strength $F_{\mu\nu}$ and $\lambda$ is the gaugino. 
Note the similarity of the structure of these terms with the operators dual to the fields $\cA_{1}$, $\cA_{0}$, which might be responsible for the presence of $C^{(2)}_{1}$ (and $C^{(2)}_{2}$, see below) in the vacuum with non-trivial baryonic current in \cite{Bigazzi:2011it}.
Let us forget about this mode in the following discussion, since it is decoupled from the rest.

We can diagonalize two other modes in the action with the change of variables
\bea
\cA_{2} &=& -\frac{Q_c \sqrt{8Q_f+\customsign^2 Q_f^2}}{\sqrt{Q_c^2+8Q_f+\customsign^2 Q_f^2}} \tilde B_{2} + \frac{\sqrt{8Q_f+\customsign^2 Q_f^2}}{\sqrt{Q_c^2+8Q_f+\customsign^2 Q_f^2}} C^{(2)}_{2} \ ,\\
{\cal B}_{2} &=& \frac{8Q_f+\customsign^2 Q_f^2}{\sqrt{Q_c^2+8Q_f+\customsign^2 Q_f^2}} \tilde B_{2} + \frac{Q_c}{\sqrt{Q_c^2+8Q_f+\customsign^2 Q_f^2}} C^{(2)}_{2}\ ,
\eea
giving the action
\bea\label{acflucads2}
S & \sim & \int  -\frac12 d\cA_{2} \wedge \sfive d\cA_{2} -\frac12 dB^{2}_{2} \wedge \sfive dB^{2}_{2}-\frac12 d{\cal B}_{2} \wedge \sfive d{\cal B}_{2} \\ 
 && \quad +\sqrt{Q_c^2+8Q_f+\customsign^2 Q_f^2} dB^{(2)}_{2} \wedge {\cal B}_{2} \ . \nonumber
\eea
Thus, upon dualization of $\cA_{2}$, we are left with a massless vector field $\tilde \cA_{1}$, dual to the dimension three flavor current.
Note that this gravity field is a combination of the quenched massless vector $\cA_{1}$ with $C^{(2)}_{1}$, $C^{(2)}_{2}$  and $B^{(2)}_{2}$.\footnote{Note that the combination requires $Q_f$-dependent dualizations and field redefinitions, so the $Q_f \rightarrow 0$ limit is not transparent at this stage.} 
Deforming the theory by the addition of (charged) flavors requires to re-define the quenched dictionary, but there is still a massless vector dual to the flavor current, which has protected dimension.

The latter is not true for other operators, as we have already seen for the $C^{(2)}_{1}$ field, since the theory is not ${\cal N}=2$.
In fact, from the rest of the action (\ref{acflucads2}) we get also a $Q_f$-dependent
 massive two-form.
In order to realize this, we dualize $B^{(2)}_{2}$ by introducing another Lagrange multiplier
\bea\label{lagrange2}
\int \tilde A_{1}\wedge d \hat H_{3}\ ,
\eea
where $\hat H_{3}\equiv dB^{(2)}_{2}$ is promoted to a basic field.
The equation of motion for $\hat H_{3}$ from (\ref{acflucads2}), (\ref{lagrange2}) gives
\bea
\hat H_{3}=-\sfive \left(d\tilde A_{1}+\sqrt{Q_c^2+8Q_f+\customsign^2 Q_f^2} {\cal B}_{2} \right)
\eea
and integrating $\hat H_{3}$ out leaves us with\footnote{We omit the part in $\cA_{2}$.}
\bea\label{acflucads3}
S & \sim & \int -\frac12 \left[ d\tilde A_{1}+\sqrt{Q_c^2+8Q_f+\customsign^2 Q_f^2} {\cal B}_{2}\right] \wedge \sfive  \left[ d\tilde A_{1}+\sqrt{Q_c^2+8Q_f+\customsign^2 Q_f^2} {\cal B}_{2}\right]\\
&& \quad -\frac12 d{\cal B}_{2} \wedge \sfive d{\cal B}_{2} \ . \nonumber
\eea
Thus, $\tilde A_{1}$ is a St\"uckelberg field and the gauge $\tilde A_{1}=0$ leaves us with the massive two-form ${\cal B}_{2}$ with mass $m^2= Q_c^2 \left(1+\frac{8Q_f+\customsign^2 Q_f^2}{Q_c^2}\right)$, and, since we work in units such that $Q_c=4$ and using $\customsign^2=1$, this corresponds to an operator of dimension $\Delta= 6+Q_f$.
In the quenched case this operator has dimension six and sits in the same multiplet of the operator dual to $C^{(2)}_{1}$ discussed above.

%
%

\subsection{Three uncharged sub-truncations}

The presence of sources in the reduced setup complicates the study of further truncations with respect to the studies in the un-sourced cases performed in \cite{Cassani:2010uw,Liu:2010sa,Gauntlett:2010vu}. There are, however, three cases in which the system simplifies consistently by turning on just a limited number of the fields presented in table \ref{tab.fields}.

\paragraph{Minimal flavored models in the Veneziano limit.} This setup consists in turning on the metric, $g$, and the scalars $f$, $w$ and $\Phi$. With this setup the flavored version of the Klebanov-Witten model and the whole infinite family of SE dual theories in the Veneziano limit was constructed  in \cite{Benini:2006hh}, and the finite temperature version was given  in \cite{Bigazzi:2009bk}. The effective action describing this truncation is
\bea
S_{KW,5d} & = & \frac{1}{2\kappa_5^2} \int \left[  R \sfive1 - \frac{40}{3}  df \w \sfive df- 20  dw \w \sfive dw- \frac{1}{2}  d\Phi \w \sfive d\Phi  - V \sfive1  \right] \ ,
\eea
with the DBI part absorbed into the potential, which we can write as
\be \label{eq.KWpotential}
V= \frac{1}{2} \left[ \frac{3}{80} \left( \frac{\partial {\cal W}}{\partial f}\right)^2 +\frac{1}{40} \left( \frac{\partial {\cal W}}{\partial w}\right)^2 +\left(  \frac{\partial {\cal W}}{\partial \Phi} \right)^2 \right] - \frac{1}{3} {\cal W}^2 \ ,
\ee
where we have defined the fake-superpotential
\be \label{eq.KWpseudopotential}
{\cal W} = e^{\frac{5}{3}f} \left[ F^{(5)}_{fl} e^{5f} + Q_f e^{f-4w+\Phi} - 4 e^{f+6w} -6 e^{f-4w}\right] \ .
\ee 
The fact that this is a consistent truncation of the (charged) D3-D7 system  implies that the analysis of hydrodynamic modes performed in \cite{Bigazzi:2009tc,Bigazzi:2010ku} is correct, since those works rely on the study of the fluctuations of a consistent truncation.

As previously mentioned, this truncation corresponds to the inclusion of vevs for a dimension eight operator of the form Tr$F^4$ and a dimension six operator of the form Tr$({\cal W}_\alpha {\cal W}^\alpha)^2$ dual to $f$ and $w$, and the insertion of the marginally irrelevant flavor term in the Lagrangian dual to $\Phi$.

\paragraph{Flavored  model with a non-trivial axion.} Turning on the five-dimensional axion, $C^{(0)}_{0}$, in the theory requires for consistency to consider the gauge fields $\AR$ and $C^{(4)}_{1}$ as well (and the related St\"uckelberg scalar $C^{(4)}_{0}$, which can be gauged away), besides the presence of the scalar fields $f$, $w$ and $\Phi$ as well as the metric $g$. The equations of motion for the axion and gauge fields read
\bea
d\left[ e^{2\Phi} \sfive F^{(1)}_{1} \right] & = & 0 \ , \\
d \left[ e^{\frac{8}{3}f-4w} \sfive F^{(5)}_{2} \right] & = & 2\sqrt{2} e^{8f+8w} \sfive F^{(5)}_{1} + F^{(5)}_{2} \w d\AR \ , \\
 d\left[ e^{-\frac{16}{3}f+8w} \sfive d \AR \right]  & = & \customsign  Q_f e^{2\Phi} \sfive F^{(1)}_{1} + e^{8f+8w} F^{(5)}_{fl} \sfive F^{(5)}_{1} + \frac{1}{2} F^{(5)}_{2} \w F^{(5)}_{2} \ .
\eea
These equations of motion can be derived from the effective action
\bea
S_{axion} & = & \frac{1}{2\kappa_5^2} \int \left[  R \sfive1-  \frac{40}{3}  df \w \sfive df- 20  dw \w \sfive dw- \frac{1}{2}  d\Phi \w \sfive d\Phi   \right. \\
&&  \qquad \qquad \left. - \frac{1}{2}  e^{2\Phi} F^{(1)}_{1} \w \sfive  F^{(1)}_{1} - \frac{1}{2}  e^{-\frac{16}{3}f+8w} d\AR\w \sfive  d\AR  \right.  \nonumber\\
&&  \qquad \qquad  \left. -  \frac{1}{2}  e^{8f+8w} F^{(5)}_{1} \w \sfive  F^{(5)}_{1} - \frac{1}{2}  e^{\frac{8}{3}f-4w} F^{(5)}_{2} \w \sfive  F^{(5)}_{2} - V \sfive1  \right] + S_{top} \ , \nonumber
\eea
with $V$ given by the expressions \eqref{eq.KWpotential}, \eqref{eq.KWpseudopotential} and the topological term being
\be
S_{top}  =  \frac{1}{2\kappa_{5}^2} \int  \frac{1}{2}   C^{(4)}_{1} \w d \AR \w dC^{(4)}_{1}   \ . 
\ee
This system describes, among others, the flavored version of the anisotropic plasma in \cite{Mateos:2011ix}, which is itself a finite temperature version of the solution in \cite{Azeyanagi:2009pr}.

In addition to the spectrum of the previous truncation we have a massless scalar (the axion, dual to the dimension four operator of the form $F \w F$) and two mixed vector fields, where the linear combinations $2\sqrt{2} C^{(4)}_{1}+F^{(5)}_{fl} \AR$ and $F^{(5)}_{fl} C^{(4)}_{1}-2\sqrt{2} \AR$ have masses $m^2=24$ and $m^2=0$ respectively, therefore corresponding to operators of dimension $\Delta=7$ and $\Delta=3$. The operator of dimension $\Delta=3$ is the surviving $U(1)_R$ gauge symmetry generator \cite{Cassani:2010uw,Gauntlett:2010vu}. Although the kinetic and mass terms of $S_{axion}$ get simplified in the new basis of vector fields, the topological term acquires a non-illuminating form, therefore we do not write it here.

\paragraph{Flavored model with brane scalar mode.} In the last consistent truncation we consider we turn on the metric $g$, the scalars $f$, $w$, $\Phi$, $B^{(2)}_{0}$ and $\cA_{0}$ and the vector field $B^{(2)}_{1}$. This system describes two different scalars coupled in St\"uckelberg fashion to the same vector, as explained in the previous sections. This is new in string theory context, since it depends crucially on the sources to give the second scalar. The relevant equations of motion for the NSNS fields are
\bea
d\left[ e^{-\Phi-\frac{4}{3}f-8w} \sfive H^{(3)}_{2} \right] & = & 2 \sqrt{2} e^{-\Phi+4(f+w)} \sfive H^{(3)}_{1} + \Theta_{4}\ ,  \\
d\left[ e^{-\Phi+4f+4w} \sfive H^{(3)}_{1} \right] & = &  e^{\Phi+\frac{28}{3}f-4w}F^{(1)}_{0}\w \sfive F^{(3)}_{0} +  e^{\frac{40}{3}f}\sfive F^{(5)}_{0} \w F^{(3)}_{0}   - \Theta_{5}\ , 
\eea
which come from the action
\bea
S_{scalar} & = &  \frac{1}{2\kappa_5^2} \int \left[  R \sfive1 - \frac{40}{3}  df \w \sfive df- 20  dw \w \sfive dw- \frac{1}{2}  d\Phi \w \sfive d\Phi \right. \\
&& \qquad\quad  \left.  - \frac{1}{2}e^{-\Phi-\frac{4}{3}f-8w} H^{(3)}_{2} \w \sfive H^{(3)}_{2}  - \frac{1}{2} e^{-\Phi+4 f + 4w} H^{(3)}_{1} \w \sfive H^{(3)}_{1}  - V \sfive1  \right] \nonumber\\
&& -\frac{4Q_f}{2\kappa_5^2} \int e^{\Phi+\frac{16}{3}f+2w} \sqrt{\left| g+ e^{-\Phi-\frac{4}{3}f-8w} \left( d\cA_{0} + B^{(2)}_{1}  \right)^{ \otimes2}\right|} \sqrt{1 + \frac{ e^{4f+4w-\Phi} }{2} \cF_{0}{^2}}\ , \nonumber
\eea
where the potential reads
\be
V  =   4 e^{\frac{16}{3}f+2w} \left( e^{10w}-6 \right) + \frac{e^{\frac{40}{3}f}}{2} \left( F^{(5)}_{fl}+\customsign \frac{Q_f}{2} \cF_{0}{^2} \right)^2  +  \frac{\customsign^2}{2} Q_f^2 e^{2\Phi + \frac{16}{3}f - 8 w}\left( 1 + e^{-\Phi+4f+4w}  \cF_{0}{^2}  \right)\ .
\ee

The scalars $\cA_{0}$ and $B^{(2)}_{0}$ and the vector $B^{(2)}_{1}$ can be decomposed into a massive vector field (with an associated St\"uckelberg scalar) and a massive scalar field $\cF_{0}$. To see this we focus on the relevant quadratic action on top of the $AdS$ background
\bea
S & = & \frac{1}{2\kappa_5^2} \int \left[- \frac{1}{2} H^{(3)}_{2} \w \sfive H^{(3)}_{2} - \frac{1}{2}H^{(3)}_{1}  \w \sfive H^{(3)}_{1}  - 2 \left( 1+2\customsign+\customsign^2 \frac{Q_f}{2}\right) \hat{\cF}_{0}{^2} \sfive 1  \right. \\
&& \qquad\qquad \left.  - \frac{1}{2} \left[ d\hat\cA_{0} + 2\sqrt{Q_f} B^{(2)}_{1}  \right] \w \sfive \left[ d\hat \cA_{0} + 2\sqrt{Q_f} B^{(2)}_{1}  \right]   \right] \nonumber \ ,
\eea
where $\hat \cA_{0}=2\sqrt{Q_f} \cA_{0}$, and $\hat \cF_{0}=\sqrt{Q_f/2}\cF_{0}$. It is a simple exercise to show that defining the scalars $\zeta=2 (\sqrt{Q_f} \cA_{0}-\sqrt{2} B^{(2)}_{0})/\mu$ and $\xi=2\sqrt{2}  \hat \cF_{0}/\mu$, with $\mu=2\sqrt{2+Q_f}$, the previous action can be written as
\bea
S & = & \frac{1}{2\kappa_5^2} \int  \left[- \frac{1}{2} dB^{(2)}_{1} \w \sfive dB^{(2)}_{1} - \frac{1}{2} \left( d\zeta+ \mu B^{(2)}_{1}  \right) \w \sfive \left( d\zeta + \mu B^{(2)}_{1}  \right)   -\frac{1}{2} d\xi \w \sfive  d\xi  \right.\\
&& \qquad \qquad \left.  - \frac{\mu^2}{4}  \left( 1+2\customsign+\customsign^2 \frac{Q_f}{2}\right)\xi^2 \sfive 1 \right] \ , \nonumber
\eea
therefore $\zeta$ is a St\"uckelberg scalar that can be gauge away, implying that $B^{(2)}_{1}$ is a vector with mass squared $m_B^{2}=\mu^2=8(1+Q_f/2)$, corresponding to an operator of dimension
\be
\Delta_B=2+\sqrt{9+4 Q_f} \ ,
\ee
and the mass of the second scalar is $m^2_\xi=4(1+Q_f/2) ( 1+2\customsign+\customsign^2 Q_f/2 )$, therefore the dimension of the associated operator depends on the sign $\customsign$, such that
\bea
\Delta_\xi = \begin{cases} 6+Q_f & {\rm if \ $\customsign=+1$}\\
2+Q_f & {\rm if \ $\customsign=-1$}
\end{cases}\ .
\eea
The $\customsign=+1$ corresponds to anti-branes, while for $\customsign=-1$ we are considering normal branes.


\section{Conclusions and future work}\label{sec.end}

In this work we have constructed the minimal consistent truncation of type IIB supergravity on squashed Sasaki-Einstein manifolds, coupled to DBI and WZ actions accounting for charged D7-branes, i.e. D7-branes with world-volume flux of the gauge field dual to the baryonic current.
We have provided the complete five-dimensional action and equations of motion, and we have discussed features of the dual spectrum of operators.

Apart from the intrinsic interest of reductions of gravity theories with explicit brane sources, our primary intention was to provide the starting point for the analysis of D3-D7 systems at finite baryon density.
In particular, the five-dimensional analysis elucidates the spectrum content of the known D3-D7 systems \cite{Bigazzi:2011it}, which was rather unclear from the ten-dimensional perspective.
Moreover, the search for new solutions and applications to quark-gluon plasma and condensed matter theories, e.g. the study of transport properties, can be undertaken starting from the five-dimensional setting.

With the equations of motion at hand, it would be very interesting to study the stability of the known charged solution - in particular, instabilities related to the world-volume gauge flux.
The consistent truncation provides the minimal consistent set of fields including the vector mode $\cA_{1}$ and the scalar $\cF_{0}$.
The former mode could be unstable towards non-homogeneous configurations at finite density, along the lines of \cite{harvey,ooguri}.

Instabilities of the charged D3-D7 system are expected on general grounds - the dual theory contains massless scalars at finite density - but are surprisingly subtle to uncover.
In particular, no instability is seen in the probe approximation \cite{Ammon:2011hz}. 
The scalar mode $\cF_{0}$ is an obvious candidate for the search of such instabilities, since without brane backreaction it sits at the edge of the BF bound (both in the uncharged and charged case).  
Thus, an obvious direction for future work is to analyze the five-dimensional equations of this paper including the fluctuation of the scalar $\cF_{0}$ on top of the backreacted background \cite{Padovaproc,inprep}.

We also pointed out novel non-baryon-charged sub-truncations of the D3-D7 systems. One of them corresponds to the flavored version of the inhomogeneous plasmas in \cite{Mateos:2011ix}.
Another one is rather new and includes the five-dimensional scalar mode coming from the brane gauge field and a single massive vector.
It would be very interesting to search for explicit (``flavored'') solutions of these systems and study the dual field theories both in the context of quark-gluon plasma and in condensed matter physics.

It would also be interesting to extend the five-dimensional reduction by including as well fields charged under the $U(1)_R$ gauge field, $\AR$. This would require the use of an extra complex two-form present in every SE manifold, as was done in the unsourced case in \cite{Cassani:2010uw,Liu:2010sa,Gauntlett:2010vu} for the matter transforming in the adjoint representation. The modifications to perform in the fields associated to the D7 worldvolume include considering two extra scalars corresponding to the transverse coordinates to each of the D7-branes and the promotion of $Q_f$ to a function of the target-manifold coordinates (such that $dQ_f\neq0$), at least in the massive flavor case.

\section*{Acknowledgments}
We are deeply indebted with Francesco Bigazzi, Davide Cassani, Stefano Cremonesi, Johanna Erdmenger, Javier Mas, Luca Martucci, Daniel Mayerson, Eric Plauschinn, Alfonso Ramallo, Alberto Santambrogio, Oscar Varela and Alberto Zaffaroni for many illuminating suggestions.
The research of A.L.C. is supported by the European Community Seventh Framework Programme FP7/2007-2013, under grant agreement n. 253937.
J.T. is supported by the Netherlands Organization for Scientific Research (NWO) under the FOM Foundation research program.

{ \it A. L. C. would like to thank the Italian students,
parents, teachers and scientists for their activity in support of
public education and research.}

\appendix

\section{Equations of motion for the NSNS sector}\label{app.NSNS}

\subsection*{Definition of the $\Theta_{n}$ forms}

The forms $\Theta_{n}$ of equations \eqref{eq.H33}-\eqref{eq.H31} are given (in coordinates basis) by
\bea
\Theta_{3} & = & - \frac{2\kappa_{10}^2}{2\cdot 3!} \frac{\varepsilon_{ab c_1 c_2 c_3}}{|\Delta_{5}|} \frac{\delta S_{DBI}}{\delta (B^{(2)}_{2})_{ab}}  dx^{c_1} \w dx^{c_2} \w dx^{c_3} \ , \\
\Theta_{4} & = & - \frac{2\kappa_{10}^2}{2 \cdot 4!} \frac{\varepsilon_{a  c_1 c_2 c_3 c_4}}{|\Delta_{5}|} \frac{\delta S_{DBI}}{\delta (B^{(2)}_{1})_{a}}  dx^{c_1} \w dx^{c_2} \w dx^{c_3}  \w dx^{c_4} \ , \\
\Theta_{5} & = & - \frac{2\kappa_{10}^2}{ 5! }  \frac{\varepsilon_{c_1 c_2 c_3 c_4 c_5}}{|\Delta_{5}|} \frac{\delta S_{DBI}}{\delta (B^{(2)}_{0})}  dx^{c_1} \w dx^{c_2} \w dx^{c_3}  \w dx^{c_4}  \w dx^{c_5} \ , 
\eea
where the $\varepsilon$ is the completely antisymmetric density tensor in five dimensions and $|\Delta_{5}|$ the volume density of the SE manifold in coordinates basis.

To understand its origin consider an action in $D$ dimensions for a $p$-form $B$ given by
\be
S = -\frac{1}{2\kappa^2} \int \frac{1}{2} dB \w \overset{D}{\star} dB + S_{s}(B) =- \frac{1}{2\kappa^2} \int d^Dx\frac{\sqrt{-G} }{2 \cdot p!} \partial_a B_{C_1 \cdots C_p} \partial^a B^{C_1 \cdots C_p}  + S_s(B_{C_1\cdots C_p}) \ ,
\ee
where $S_s$ is a source term, which in the simplest DBI case is given roughly by $S_s \sim \int \sqrt{-( G+  B)}$. The equations of motion for the components $B_{C_1\cdots C_p}$ of the $p$-form are given by the Euler-Lagrange equations 
\be
 \partial_A \left( \sqrt{-G} \partial^A B^{C_1 \cdots C_p}\right) + 2\kappa^2 \frac{\delta S_s}{\delta B_{C_1 \cdots C_p}} =0\ .
\ee
We focus now on the derivative term, we manipulate it in the following way
\be\label{componentstoform}
\frac{\varepsilon_{C_1 \cdots C_p A_1 \cdots A_{D-p}}}{p! (D-p)!}   \partial_A \left( \sqrt{-G} \partial^A B^{C_1 \cdots C_p}\right) dX^{A_1} \w \cdots \w dX^{A_{D-p}} =   \overset{D}{\star} \delta\, dB  = (-1)^{D+p^2}d \left( \overset{D}{\star}  dB \right)  \ ,
\ee
where $\delta$ is the codifferential.\footnote{The codifferential acting on a $q$-form is given by $\delta= (-1)^{D(q+1)+1} \overset{D}{\star}d\overset{D}{\star}$, which together with the identity in Lorentzian signature $\overset{D}{\star} \overset{D}{\star} F_{p} = (-1)^{p(D-p)+1} F_{p}$, gives the second equality in \eqref{componentstoform}.} Performing the same manipulation to the whole equations of motion, in the case with $D=10$ and $p=2$, we have (reinstating a factor of the dilaton)
\be\label{eq.eomHappendix}
d\left ( e^{-\Phi} \sten dB_{2} \right) + \frac{2\kappa_{10}^2}{2! \, 8!} \varepsilon_{C_1  C_2 A_1 \cdots A_8}  \frac{\delta S_{DBI}}{\delta B_{C_1  C_2}} dX^{A_1} \w \cdots \w dX^{A_{8}}  + \text{extra terms} = 0\ ,
\ee
where the extra terms come from  pieces of the action we have not considered above. Upon reduction, the second term in \eqref{eq.eomHappendix} gives rise to the $\Theta_{n}$ terms written in (\ref{theta3})-(\ref{theta5})  (plus a sign since in \eqref{eq.H33}-\eqref{eq.H31} these terms are on the right-hand side).

\subsection*{Relation to the equations of motion for the gauge field $\cA$}

Here we show how equations \eqref{eq.H33}-\eqref{eq.H31} are  independent, and their exterior derivatives give the equations of motion for $\cA_{0}$ and $\cA_{1}$. Before doing this for the five-dimensional reduced fields, let us study first what happens in the ten-dimensional case. The following argument will rely on the equations of motion derived from the action \eqref{eq.oldaction}, but since this system and \eqref{eq.newaction} give rise to the same dynamics, the conclusion can be used in our reduction to five dimensions, which is based on the latter to impose the self-duality of the $F_{5}$ field strength in a neat form.

In \eqref{eq.oldaction} the world-volume gauge field comes always in the combination $\cF = 2\pi \alpha' d\cA+B_{2}$ and consistency with $d^2=0$ implies that the equation of motion for the gauge field, $\cA$, follows from the equation of motion for the NSNS potential $B_{2}$
\be
\text{EOM}[\cA] = d \frac{\delta S}{\delta d\cA} = 2\pi \alpha' d \frac{\delta S}{\delta \cal F}= 2\pi \alpha' d \left( \frac{\delta S}{\delta  B_{2}} + d \frac{\delta S}{\delta  dB_{2}}\right) =  2\pi \alpha' d \left(  \text{EOM}[B_{2}] \right)\ .
\ee

In the reduced case the NSNS potential appears in two different ways: through $H^{(3)}_{1,2,3}$ in the IIB part of the action and through $\cF_{0,1,2} = 2\pi\alpha' F_{0,1,2}+B^{(2)}_{0,1,2}$ in the sources part of the action.\footnote{Notice that the topological term is written as $\int H_3 \wedge F_3 \wedge C_4$.} Therefore, the equation of motion for $B^{(2)}_{2}$, $B^{(2)}_{1}$ and $B^{(2)}_{0}$ are given respectively by
\bea
d \frac{\delta S_{IIB}}{\delta H^{(3)}_{3}} - \frac{\delta S_{sources}}{\delta \cF_{2}} & = & 0 \ , \label{eomB2} \\
d \frac{\delta S_{IIB}}{\delta H^{(3)}_{2}} - 2\sqrt{2} \frac{\delta S_{IIB}}{\delta H^{(3)}_{1}} + \frac{\delta S_{sources}}{\delta \cF_{1}} & = & 0 \ ,  \label{eomB1} \\
d \frac{\delta S_{IIB}}{\delta H^{(3)}_{1}} - \frac{\delta S_{sources}}{\delta \cF_{0}} & = & 0 \ . \label{eomB0} 
\eea
On the other hand, the equations of motion for $\cA_{1,0}$ are obtained just from their presence in $S_{sources}$ via $\cF$ as
\bea
d \frac{\delta S_{sources}}{\delta \cF_{2}} & = & 0 \ ,  \label{eomA1} \\
d \frac{\delta S_{sources}}{\delta \cF_{1}} - 2\sqrt{2}\frac{\delta S_{sources}}{\delta \cF_{0}} & = & 0 \ . \label{eomA0} 
\eea
Acting with the exterior derivative in \eqref{eomB2} gives the equation of motion \eqref{eomA1}. However, acting with $d$ on \eqref{eomB1} gives the combination
\be
d\, \text{EOM}[\eqref{eomB1}] = 2\sqrt{2} \, \text{EOM}[\eqref{eomB0}] - \text{EOM}[\eqref{eomA0}] \ .
\ee
Finally, acting with $d$ on \eqref{eomB0} gives zero identically.

\section{Reduction of the five-form and topological terms}\label{app.topological}

Starting from the original action \eqref{eq.oldaction},  to reduce the five-form and impose its self-duality relations the topological terms must be written in terms of $F_{5}$ instead of $C_{4}$. With the definitions \eqref{eq.newdefF1}-\eqref{eq.newdefF5} it is straightforward to check that the equations of motion in ten dimensions are correctly derived from the action \eqref{eq.newaction}. To achieve the reduction, we proceed by modify minimally the approach in \cite{Cassani:2010uw}. Let us  reduce the five-form plus topological term pieces. We introduce the form
\be
L_5 \equiv B_2 \w dC_2 - C_2 \w dB_2 + \cF^2 \w F_1^{D7} \ .
\ee
Note that $d(F_{5} -\frac{1}{2} L_{5})=0$.
The decomposition in five-dimensional fields is given in (\ref{L5bis})-(\ref{L5bisend}).

The reduction of the five-form piece of the action, prior to imposition of the self-duality, gives
\be \label{fivereduced}
-\frac14 \int_{{\cal M}_{10}}  F_{5} \w \sten F_{5} \quad \longrightarrow \quad -\frac14 V(X_{SE})  \int_{{\cal M}_5} \sum_{n=0}^5 \Gamma_{n,5-n} F_{n}^5 \w \sfive F_{n}^5 \ ,
\ee
with $\Gamma_{n,5-n}$ defined in \eqref{eq.Gammapn}.

Analogously, the first part of the topological term (\ref{topo}) is simply reduced as
\be\label{topo1}
\frac14 \int_{{\cal M}_{10}}(F_{fl}^{(5)} + dC_{4}) \w L_{5} = \frac{1}{4}\int_{{\cal M}_{10}} F_{5} \w L_{5}  \quad \longrightarrow \quad \frac{1}{4} V(X_{SE})  \int_{{\cal M}_{5}}\sum_{n=0}^5 (-1)^{n+1} F_{n}^{(5)} \w  L_{5-n}^{(5)} \ .
\ee
Before reducing the other pieces of the topological term, let us impose the self-duality of the five-form.
This can be achieved, following \cite{Cassani:2010uw}, by adding to the action the Lagrange multiplier 
\bea\label{multiplier}
{\cal L}_{mult} & = & \frac{V(X_{SE})}{4\kappa_{10}^2}  \Bigg[ \left(F_{5}^{(5)}-\frac12 L^{(5)}_{5} \right) F^{(5)}_{fl} - \left(F_{4}^{(5)}-\frac12 L^{(5)}_{4} \right) \w \left( dC^{(4)}_{0} -2 \sqrt{2} C^{(4)}_{1} - F^{(5)}_{fl} \AR \right) \nonumber \\
&& \quad \quad  + \left(F^{(5)}_{3}-\frac12 L^{(5)}_{3} + C^{(4)}_{1} \w d\AR \right) \w dC^{(4)}_{1} \Bigg] \ .
\eea
It can be checked that varying the whole action (\ref{fivereduced}-\ref{multiplier}) w.r.t. $F_{5}^{(5)}, F_{4}^{(5)}, F^{(5)}_{{3}}$ the self-duality conditions follow, and that varying it w.r.t. $C^{(4)}_{0},C^{(4)}_{1}$ the Bianchi identities follow.

Now we can substitute the self-duality relations back into the action, getting rid of $F_{5}^{(5)}, F_{4}^{(5)}$ and $F^{(5)}_{3}$.
This substitution kills the term (\ref{fivereduced}), while the terms (\ref{topo1}), (\ref{multiplier}) reduce to the standard kinetic term
\be
S_{kin,mult}=\frac{1}{2\kappa_{5}^2} \int \frac{-1 }{2} \sum_{n=0}^2 \Gamma_{n,5-n}  F_{n}^{(5)} \w \sfive F_{n}^{(5)} \ , 
\ee
and a topological term
\bea \label{topo2}
S_{top,mult}& = &\frac{1}{2\kappa_{5}^2} \int \frac{1}{4}  \Bigl[ - \left(F^{(5)}_{0}+F^{(5)}_{fl} \right)\w L^{(5)}_{5}  + \left(F^{(5)}_{1}+dC^{(4)}_{0} - 2\sqrt{2} C^{(4)}_{1} -F^{(5)}_{fl}\AR  \right)\w L^{(5)}_{4}  \nonumber \\
&& \qquad \qquad \quad -\left(F^{(5)}_{2}+dC^{(4)}_{1} \right)\w L^{(5)}_{3} + 2 C^{(4)}_{1} \w d \AR \w dC^{(4)}_{1} \Bigr]\ . 
\eea

Now we can reduce the rest of the topological term (\ref{topo}), by re-expressing it as
\be
S_{top,rest} =  \frac{1}{8}  L_{5}  \w \cF^{2} \w F_{1}^{D7}- \frac{1}{6} F_{3} \w \cF^{3} \w F_1^{D7} + \frac{1}{24} F_{1} \w \cF^4 \w F_1^{D7} \ ,
\ee
which reduces in a straightforward manner to
\bea\label{topo3}
S_{top,rest} & = &\frac{1}{2\kappa_{5}^2}  \int \frac{ \customsign \,Q_f}{2}  \Bigg[  { \cF_{0}}^2
 \left[ \frac{1}{4}   L^{(5)}_{5}   - \frac{1}{3}  F^{(3)}_{3} \w \cF_{2}  + \frac{1}{6}  F^{(1)}_{1} \w {\cF_{2}}^2  \right] \ \\
 && \qquad +{ \cF_{0}}^2 \Bigg[ \frac{1}{4}   L^{(5)}_{4}   - \frac{1}{3}  \left(    F^{(3)}_{3} \w \cF_{1} +   F^{(3)}_{2} \w \cF_{2}  \right)  \nonumber \\
 && \qquad  \qquad  \qquad  + \frac{1}{12}  \left(   2 F^{(1)}_{1} \w \cF_{2} \w \cF_{1} + F^{(1)}_{0} \w {\cF_{2}}^2 \right)    \Bigg] \w \AR  \nonumber \\
 && \qquad + 2 \cF_{0}  \left[ \frac{1}{4}   L^{(5)}_{3}   - \frac{1}{3}  \left(    F^{(3)}_{3} \w \cF_{0} +   F^{(3)}_{1} \w \cF_{2}  \right)  \right.  \nonumber \\
 && \qquad  \qquad  \qquad \left.  + \frac{1}{6}  \cF_{0} F^{(1)}_{1} \w \cF_{2}  \right] \w \left(\cF_{2} + \cF_{1} \w \AR \right)  \nonumber \\
 && \qquad + 2 \cF_{0}  \Bigg[ \frac{1}{4}   L^{(5)}_{2}   - \frac{1}{3}  \left(    F^{(3)}_{2} \w \cF_{0} +   F^{(3)}_{1} \w \cF_{1} +   F^{(3)}_{0} \w \cF_{2}  \right) \nonumber \\
 && \qquad \qquad \qquad + \frac{1}{6} \cF_{0}  \left(  F^{(1)}_{1} \w  \cF_{1} + F^{(1)}_{0} \w \cF_{2} \right)    \Bigg]  \w \cF_{2} \w \AR \nonumber \\
 && \qquad +  \cF_{2} \w \left(\cF_{2} +2 \cF_{1} \w \AR \right)  \w \left[ \frac{1}{4}   L^{(5)}_{1}   - \frac{1}{3} \cF_{0}  F^{(3)}_{1}     + \frac{1}{12}  {\cF_{0}}^2   F^{(1)}_{1}    \right] \nonumber \\
 && \qquad  + {\cF_{2}}^2 \w \AR  \w \left[ \frac{1}{4}   L^{(5)}_{0}   - \frac{1}{3} \cF_{0}  F^{(3)}_{0}     + \frac{1}{12}  {\cF_{0}}^2   F^{(1)}_{0}    \right]  \nonumber \Bigg]\ .
\eea

Adding up \eqref{topo2} and \eqref{topo3} we arrive to the result in \eqref{eq.topological} after some mild manipulations.

\section{Reduction of the smeared DBI term}\label{app.DBI}

For each D7-brane we have an embedding profile and an induced metric on the worldvolume of the D7 given by the pullback
\be
\hat G_{ab}=\frac{\partial X^M}{\partial \xi^a}\frac{\partial X^N}{\partial \xi^b} G_{MN} \ ,
\ee
with $\hat G$ the pullback metric, $G$ the ten-dimensional metric, $M,N=0,\cdots,9$, and $a,b=0,\cdots,7$. The variation of the DBI action with respect to the ten-dimensional metric components can be written accordingly as
\be
\frac{\delta S_{DBI}}{\delta G_{AB}} = \frac{\delta S_{DBI}}{\delta \hat G_{ab}} \frac{\partial X^A}{\partial \xi^a}\frac{\partial X^B}{\partial \xi^b} \ .
\ee
We consider only trivial embeddings in this paper.

The smearing procedure affects the DBI action in the following way
\be\label{smearing}
S_{DBI} = -\frac{1}{2\kappa_{10}^2} \sum^{N_f} \int d^8 \xi e^\Phi \sqrt{-\det \left( \hat G + e^{-\frac{\Phi}{2}}\cF \right)} \rightarrow -\frac{4Q_f}{2\kappa_{10}^2} \int d^{10} X e^\Phi \sqrt{-\det \left( \hat  G + e^{-\frac{\Phi}{2}}\cF \right) } \ ,
\ee
implying in the static gauge
\be
\frac{\delta S_{DBI}}{\delta G_{AB}}  = - \frac{2Q_f}{2\kappa_{10}^2} e^\Phi \sqrt{-\det \left( \hat G + e^{-\frac{\Phi}{2}}\cF \right) }  \left( \left(  \hat G + e^{-\frac{\Phi}{2}}\cF \right)^{-1} \right)^{(AB)} \ .
\ee

To write the five-dimensional equivalent to this term we need to translate $ \sqrt{-\det \left( \hat  G + e^{-\frac{\Phi}{2}}\cF \right) } $ to five-dimensional language. After a quick calculation, using
\be
\det \begin{pmatrix} A & B \\ C & D \end{pmatrix} = \det D\, \det\left( A-BD^{-1}C \right) \ ,
\ee
 one arrives to the answer
\be
\sqrt{-\det \left( \hat  G+e^{-\frac{\Phi}{2}} \cF \right)}  =  e^{\frac{16}{3}f+2w} \sqrt{1+\frac{e^{4f+4w-\Phi}}{2}\cF_{0}^2} \sqrt{-\det(Z)} |\Delta_{5}|\ ,
\ee
where the matrix $Z$ is given by
\be
Z=g+ e^{-\frac{\Phi}{2}-\frac{10}{3}f} \left( \cF_{2} + \cF_{1} \w \AR \right)  + e^{-\Phi-\frac{4}{3}f-8w}\cF_{1}\otimes \cF_{1}   \ ,
\ee
and we have been careful in not over-counting the $\left( 1+\frac{e^{4f+4w-\Phi}}{2} {\cF_{0}}^2 \right)$ factor coming from the the two directions transverse to the D-branes. In other words, since in the smearing procedure we promote the eight-dimensional world-volume to a ten-dimensional factor, we have to quotient w.r.t. the full open string two-dimensional metric gained in the procedure; the quotient brings an extra $1/\sqrt{1+\frac{e^{4f+4w-\Phi}}{2} {\cF_{0}}^2}$ factor on the r.h.s. of (\ref{smearing}).

Therefore, after integrating out the SE space, the DBI term may be written as
\be
S_{DBI} = -\frac{4Q_f}{2\kappa_{5}^2} \int e^{\Phi+\frac{16}{3}f+2w}\sqrt{\det(g^{-1}\cdot Z)} \sqrt{1+\frac{e^{4f+4w-\Phi}}{2}   {\cF_{0}}^2} \sfive 1 \ .
\ee

The contribution of the DBI term to the equation of motion for $\AR$ is given by
\be
\frac{ \delta S_{DBI} }{ \delta A_{1} }= d \left( \frac{ \hat\delta S_{DBI} }{ \hat\delta dA_{1} } \right) +\frac{ \hat\delta S_{DBI} }{ \hat\delta A_{1} } \ ,
\ee
where the hatted variation is taken considering $A_{1}$ and $dA_{1}$ as independent fields. This calculation yields 
\bea
\frac{\hat \delta S_{DBI} }{\hat \delta dA_{1} } & = & \frac{\hat \delta S_{DBI} }{ \hat\delta \cF_{2} } \w \frac{ \hat\delta \cF_{2} }{ \hat\delta dA_{1} } = \Theta_{3} \cA_{0} \ , \\
\frac{\hat \delta S_{DBI} }{ \hat\delta A_{1} } & = & \frac{ \hat\delta S_{DBI} }{ \hat\delta (\cF_{2} + \cF_{1} \w A_{1}) } \w \frac{ \hat\delta (\cF_{2} + \cF_{1} \w A_{1}) }{ \hat\delta A_{1} } = \frac{ \hat\delta S_{DBI} }{ \hat\delta \cF_{2}  } \w \cF_{1} = \Theta_{3} \w \cF_{1} \ ,
\eea
so the total variation is the one written in \eqref{eq.U1Reom}.

\end{document}